\documentclass[%
 aip,
 amsmath,amssymb,
 reprint,longbibliography,
]{revtex4-1}
\usepackage[utf8]{inputenc}
\usepackage{graphicx}
\usepackage{bbold}
\usepackage{cases}
\usepackage{upgreek}
\usepackage{bm}
\usepackage[usenames, dvipsnames]{color}
\usepackage[colorlinks=true, citecolor=RedOrange, linkcolor=MidnightBlue, urlcolor=BlueViolet ]{hyperref}
\usepackage{braket}
\usepackage[normalem]{ulem}
\usepackage{xfrac}
\usepackage{subcaption} 
\usepackage{ragged2e} 
\DeclareCaptionJustification{justified}{\justifying}

\usepackage{textgreek}
\usepackage{gensymb}
\usepackage{dcolumn}

\newcommand \mc[1] { \mathcal{#1} }
\newcommand \dd[1]  { \!\textrm d{#1} \,}   
\newcommand \rmm[1]  { \textrm{#1} }

\newcommand \e[1] { \rmm{e}^{#1} }

\newcommand \mel[3] { \left\langle #1 \right| #2 \left|  #3 \right\rangle  }

\newcommand \imi { \mathrm{i}}

\makeatletter
\def\@email#1#2{%
 \endgroup
 \patchcmd{\titleblock@produce}
  {\frontmatter@RRAPformat}
  {\frontmatter@RRAPformat{\produce@RRAP{*#1\href{mailto:#2}{#2}}}\frontmatter@RRAPformat}
  {}{}
}%
\makeatother
\begin{document}

\preprint{AIP/123-QED}

\title{Efficient spectra from atomistic simulation: a generalized master equation study of the air-water interface}

\author{Thomas Sayer}
\affiliation{Department of Chemistry, Durham University, Durham DH1 3LE, United Kingdom\looseness=-1} 


\date{\today}

\begin{abstract}
Computing condensed phase spectra from atomistic simulations requires calculating correlation functions from molecular dynamics and can be very expensive. A totally general, data-driven method to reduce cost is to employ an exact rewriting to a generalized master equation characterized by a memory kernel. The decay time of the kernel can be less than the original function, reducing the amount of data required, but it can also be more. In this paper we construct the minimal projection operator to predict vibrational sum-frequency generation spectra and apply it to the air-water interface simulated using ab initio molecular dynamics. We find the kernel is shorter lived than the correlation functions, yielding equivalent spectra when truncated to around 50\% the duration. We explore various avenues to use more of the available data to expand the projector in an attempt to reduce the cost further. Interestingly, we are not able to effect any change by including quadrupoles, inter-molecular couplings, or a depth-dependence. How to strategically go about maximally reducing cost using projection operators remains an open question.
\end{abstract}

\maketitle

\section{Introduction}
\vspace{-6pt}

It is a central goal of chemical physics to link theoretically calculated microscopic dynamics with experimentally measured macroscopic observables, perhaps the most important being optical spectra. From the simulation side, one of the persistent challenges for molecular dynamics approaches in accessing macroscopic quantities is that of computational cost. New algorithms to evolve the equation of motion (EOM) address the enduring objective of extending the feasible computational timescale for systems of a given size, whether the forces be coarse-grained,\cite{Stevens2023, Gaudy2024} atomistic forcefield (FFMD),\cite{BookFrenkelSmit} ab-initio Born-Oppenheimer\cite{Liberatore2018} (AIMD) or non-adiabatic,\cite{Andermatt2016} ring-polymer quantum analogue,\cite{Lawrence2023} or otherwise. Indeed, one of the guiding principles in choosing a particular level of the computational hierarchy is that the characteristic timescales of the phenomenon of interest can be captured within the window available. For condensed phase spectroscopy, inherently quantum observables like the transition dipole moment must be measured over correlation times involving heavier, nuclear degrees of freedom that we might otherwise want to treat as classical, creating a tension between what is desired and what is affordable.

This problem of affordable timescale is intrinsic to the study of quantum processes in condensed media, and so there are a wealth of strategies available with which to attack it. As one example, polaron forming systems that exhibit coupling between electronic and quantum-nuclear motion can exhibit dynamics over many 10s or 100s of picoseconds,\cite{Ginsberg2020} far beyond what is typically accessible to fully quantum dynamical approaches. One approach to bridge regimes is to move to new methods that trade accuracy (and sometimes clarity) for efficiency: in this example, semiclassical approaches like surface hopping\cite{Giannini2022} or embedding methods such as QM/MM.\cite{Burke2024} Another line of attack is to map to a model system that reduces the dimensionality of the problem whilst being parameterized by data at the higher level of theory or experiment.\cite{Bhattacharyya2024a} A~related route is to parameterize the forces (or other observables, like the total current\cite{Bhattacharyya2024b}) directly using machine learning, particularly neural network potentials that can, for example, allow AIMD accuracy at FFMD cost with a relatively small set of reference calculations.\cite{Libbi2025} In this paper we employ the generalized master equation (GME), which is a model-free way to compute correlation functions at possibly much reduced cost. Here, we will use only the natural outputs of AIMD simulation. However, our approach to the method is completely date-driven, meaning the dynamics could in principle be obtained at any level of theory, with any of the aforementioned methods, and it can even be extended to treat spatial finite-size effects.\cite{Bhattacharyya2025a}

The GME describes the coupled EOMs for a set of correlation functions.\cite{BookChapterBerne} It is obtained in three steps: by choosing some `observables'\footnote{
Strictly speaking, the dipole moment in a periodic system is not an observable.\cite{Resta1998} Only changes in the total polarization are well-defined. Nevertheless the operator we associate with the dipole (molecular or total) can still be written and computed within the limits of the Berry-phase polarization branch,\cite{Gaigeot2003} provided the system remains insulating\cite{Spaldin2012}
} 
$|\bm{A})$, using the projection operator $\mc{P}=|\bm{A})(\bm{A}|\bm{A})^{-1}(\bm{A}|$ to define a generalized Langevin equation of motion for some (in principle different) observables~$|\bm{B})$, and then forming correlation functions by taking the inner product with a third set of observables~$|\bm{D})$. When all observables are contained within $|\bm{A})$, the expression\cite{Nakajima1958a, Zwanzig1960, Mori1965b}
\begin{equation}\label{eq:MNZ}
    \dot{\bm{\mc{C}}}(t) = \dot{\bm{\mc{C}}}(0)\bm{\mc{C}}(t)  - \int_0^t \dd{s} \bm{\mc{K}}(s)\bm{\mc{C}}(t-s)
\end{equation}
is a formally exact, non-Markovian EOM for the matrix of correlation functions $\bm{\mc{C}}(t)\equiv(\bm{A}(0)|\bm{A}(t))$. The observables could be classical or quantum mechanical.\cite{BookChapterBerne} The dynamics of $\bm{\mc{C}}(t)$ are completely determined by its initial condition $\dot{\bm{\mc{C}}}(0)$ and the memory kernel, $\bm{\mc{K}}(t)$. Efficiency gains are possible when $\bm{\mc{K}}(t)$ is a decaying function with some characteristic lifetime $\tau$ that is less than the times of $\bm{\mc{C}}(t)$ we wish to reach. Since $\bm{\mc{K}}(t')$ can be extracted from $\bm{\mc{C}}(t\leq t')$ by numerical inversion,\cite{BookBoonYip} accurate time series data for $\bm{\mc{C}}(t)$ up to $\tau$ allow access to $\bm{\mc{C}}(t>\tau)$ at no reduced accuracy and for (comparatively) trivial cost. In the best case scenario the matrix $\bm{\mc{C}}(t)$ is generated by a Markov process and $\bm{\mc{K}}(t)\sim \delta(t)$: the exponential decays of $\bm{\mc{C}}(t)$ (with some longest lifetime which can be arbitrarily large depending on the physics of the system) are completely described by the first time step $\bm{\mc{C}}(\Delta t)$.\cite{Husic2018} In many practical cases in both quantum dynamics\cite{Pfalzgraff2019a, Bhattacharyya2024b, Sayer2024, Trenins2025}$^,$\footnote{The authors of Ref.~\onlinecite{Trenins2025} have informed us the lifetime of the RPMD kernel is just 25~fs, two orders of magnitude shorter than the transmission coefficient plateau time in the low-friction regime.} and biophysics,\cite{Cao2020a, Dominic2022, Cao2023} a large separation of timescales has allowed cost reduction of many orders of magnitude.

Unfortunately for the applicability of the GME method in computational spectroscopy, memory kernels extracted from atomistic MD are often found to be as long (or even much longer) lived than the timescales of the correlation functions they parametrise.\cite{Lange2006, Carof2014a, Lesnicki2016, Kowalik2019, Bruenig2022, Vroylandt2022a, Vroylandt2022b, KieferArchive} Indeed, this computationally unfavourable regime $\tau > t_\rmm{eq}$ arises in applications of the GLE more generally, e.g. to time series describing local weather conditions.\cite{Kiefer2025} The reason for this is that the memory kernel's lifetime is a measure of the slowest timescale within the complementary space of ${\mc{P}}$, and so acceleration of computation using a non-Markovian EOM to bypass full propagation can only\footnote{
There is a case where the memory kernel fits a model form that can be extrapolated,\cite{Klimek2025} but in general long-timescales in the memory kernel cannot easily be inferred from short-time data, especially in the presence of noise.
} 
be successful if the projection operator contains both the slowest timescales of the system and the observable of interest. The observable of interest does not \textit{itself} have to be the slowest timescale. Hence, a projection that includes many degrees of freedom could result in an efficiency gain while a more minimal, obvious choice may lead to none. Ultimately, the definition of the projection operator is a choice made by the practitioner, informed by physical intuition and experience of what the dynamical eigenspectrum might look like.\cite{BookChapterBerne}$^,$\footnote{
This can be seen as the opposite of a method like DMD that takes the full configurational space and extracts the high/infinite dimensional (Markovian) propagator by performing a dimensionality reduction, effectively approaching the minimal space from above by rank reduction.\cite{Schmid2022}
} 
The central question this study will address is therefore, ``Do~GMEs constructed with spectroscopic correlation functions from AIMD calculations have short-lived memory kernels and, if not, what can be done to amend the projection operator so that they do?''

In what follows, we will look to employ the GME approach in calculating vibrational sum-frequency generation (VSFG) spectra from AIMD simulations. To keep things simple, computationally, and because it has been extensively studied,\cite{Gaigeot2003, Auer2008, Stiopkin2011, Sulpizi2012, Nagata2013, Ohto2015, Khatib2016, Khatib2017, Moberg2018, Litman2023, Jana2024, Fellows2024, Fellows2024b} we will take our system to be the air-water interface. The signals we seek to predict are particular directional combinations of the polarizability-dipole correlation,\cite{BookMorita, BookShen}
\begin{equation}\label{eq:corr_2nd_vsfg}
    \chi^{(2)}_{ijk}(\omega) \simeq X(\omega) \int \dd{t} \e{\imi \omega t} \langle \alpha_{ij}(0)\mu_k (t) \rangle,
\end{equation}
where $\alpha_{ij}(\bm{R}) = \Delta\mu_i(\bm{R}) / E_j$ is computed as a finite difference of dipole moments in the presence and absence of a small external electric field that serves to induce a distortion in the density $\rho(\bm{r};\bm{R})$ for fixed nuclei $\bm{R}$.\cite{Sulpizi2012} This amounts to taking the resonant term in the full expression for the second order response (see Appendix~\ref{app:SFG_derivation}). At this level of theory the nuclei are taken to be classical, which is justified by replacing the exact Kubo-transformed correlation function with its classical counterpart, choosing prefactor to be $X(\omega)=\imi\beta\omega/2$.\cite{Morita2006} Therefore we can use Born-Oppenheimer based MD with explicit electronic density to calculate the correlation functions. 
\vfill
\section{Methods}\label{sec:methods}
\vspace{-8pt}
We pre-requilibrated a $17\times17$~\AA~slab of $171$~waters using SPC/E\cite{SPCE} in GROMACS,\cite{GROMACS} the $z$~direction was taken as $L=57$~\AA~to avoid image interactions in the Ewald sum.\cite{BookFrenkelSmit} We then ran AIMD\cite{VandeVondele:2005ge, VandeVondele:2003ue} with the PBE functional\cite{Perdew1996} plus D3 vdW correction\cite{GrimmeD3} using CP2K\cite{CP2K} at $350$~K. We ran $2000$ steps of $0.5$~fs under massive thermostat to re-equilibrate and then $104,000$~steps under standard CSVR\cite{Bussi2007} with a time constant of $100$~fs. It is known that good convergence can take upwards of 3~ns,\cite{Ohto2015} so our results are expected to be inaccurate and should be understood as an illustration of the GME method. If we find we are in beneficial regime, converged spectra could be obtained even though the reference dynamics are themselves underconverged (see Appendix~\ref{app:cost_reduction}). The finest plane wave cutoff was $350$~Ry, with $4$~grids used at a relative grid ratio of $40$~Ry. The basis set was DZVP-MOLOPT-SR-GTH with the GTH pseudopotential.\cite{Goedecker:1996ve, Hartwigsen1998} We post-processed the trajectory by printing the electron density every $8$~time steps -- chosen as the maximum stride that would still allow $\omega_\rmm{max}$ to include the O--H stretching region -- as well as under an imposed electric field of $0.005$~a.u. in the $x$, $y$, and $z$~directions separately.

The four time series were input to the open-source TRAVIS code\cite{TRAVIS} for radical Voronoi partitioning,\cite{Rycroft2009} which calculated atomic contributions to the electronic density and dipole (and quadrupole);\cite{Thomas2015} units are e and e~pm respectively. TRAVIS output files report atomic dipoles $\int_\mathrm{atom}\dd{\bm{r}}\rho(\bm{r})(\bm{r}-\bm{r}_\mathrm{atom})$, whilst the total dipole is
\begin{equation}\label{eq:travis_mu}
    \bm{\mu} = \sum_\mathrm{mols}\sum_\mathrm{atoms \in mol}\int_\mathrm{atom}\hspace{-10pt}\dd{\bm{r}}\rho(\bm{r})(\bm{r}-\bm{r}_\mathrm{origin}).
\end{equation}
Since atomic cells bore net charge (oxygen around $-1.4$~e and hydrogen around $0.7$~e), when values were combined into molecules their origins were re-referenced $(\bm{r}_\mathrm{atom}-\bm{r}_\mathrm{mol})\int_\mathrm{atom}\dd{\bm{r}}\rho(\bm{r})\equiv (\bm{r}_\mathrm{atom}-\bm{r}_\mathrm{mol})q_\mathrm{atom}$, where we chose $\bm{r}_\mathrm{mol}$ as the common centre of mass. The same referencing step then applies moving from $\bm{r}_\mathrm{mol}$ to $\bm{r}_\mathrm{origin}$, which we chose as the slab centre of mass since this removes any drift in the center of mass. A similar procedure was also applied to the quadrupoles.

The standard routine in TRAVIS returns the spectrum appropriate to completely neutral molecules. However molecules were only on-average neutral, with FWHM of around $0.04$~e. The electric dipole of two charged water molecules with a typical distance of 3~\AA~and a charge of $\pm 0.04$~e is approximately $0.6$~D,\footnote{We would like to thank the first anonymous reviewer for pointing out the significance of neglecting the net molecular charges.} therefore not referencing each molecule to the common origin discards a substantial charge current. Indeed the IR spectrum with neutral molecules has the wrong band intensity ratios: it suggests all bands are roughly equally intense, but the stretching region should be around 3~times as intense, experimentally.\cite{Bruenig2022, Bertie1996} We performed all the following analysis with and without taking into account the molecular charges and, while all spectra changed significantly, the conclusions about the functionality of the GME were the same. Below, we present the figures appropriate to the re-referenced case, as they bare a much closer resemblance to experiment. 

Polarizabilities were computed as finite differences between the dipoles calculated with and without the applied external field. To correctly account for the inversion symmetry in our slab system, molecular $\alpha_{ij}$ were passed through a switching function that accounts for the surface normal and also discards contributions close to the central slab layer to reduce noise, as in Ref.~\onlinecite{Litman2023}.

From these time series data we computed the correlation functions, averaging windows of length $1024$~steps with half-overlapping steps of length~$512$. We always worked with time-derivatives of the observables and analytically pre-multiplied by $1/(\rmm{i}\omega)^2$.\cite{Ohto2015} Cross terms between molecules were spatially truncated\cite{Nagata2010, Nagata2013} at $5$~\AA\cite{Ohto2015} using a neighbour list constructed at the start of the averaging window and appropriate for periodic boundaries as implemented in \textit{scipy.spatial.cKDTree}. When Fourier transforming to give the spectra the functions were zero-padded by a factor~$3$, mirrored, and windowed using a Hann function; the sinc correction for finite time step was applied to get the correct ratio of band intensities. When filtering the raw MD data, a piecewise function that interpolates between $1$ and $0$ using the Hann window over 43~steps was employed, the longest we could use to have a comparison to our GME result; the longer the tapering function, the less artificial ringing is introduced. If shorter windows are used the shoulder in $\chi^{(2)}_{\parallel \parallel z}$ disappears.

For robustness to noise, we employed the derivative-free TTM~\cite{Cerrillo2014a} numerical approach to the non-Markovian equation of motion,
\begin{equation}\label{eq:TTM}
    \bm{\mc{C}}_n= \sum_{i=1}^{\mathrm{max}(n,\tau)} \bm{\mc{T}}_{i}\bm{\mc{C}}_{n-i},
\end{equation}
where $\bm{\mc{T}}_t$ is obtained iteratively, equivalent to how the memory kernel is often obtained in the GLE\cite{Kiefer2025} (but without any trapezium rule weights for the endpoints). The extracted memory object $\bm{\mc{T}}_t$ equals $\bm{\mc{K}}(t)$ in the limit ${\Delta t\rightarrow 0}$. We find these to be in close quantitative agreement and full qualitative agreement with the memory kernels extracted using traditional Volterra methods when comparison is possible. To not unduly complicate the paper, we refer to the memory object as the kernel, or $\bm{\mc{K}}(t)$, below. To evolve the correlation function in time, the convolution in Eq.~\ref{eq:TTM} is computed iteratively, which by construction returns the original series when no cutoff is used. When a cutoff is used, the upper limit on the sum is replaced by $\tau~\forall~n > \tau$, which means that only data from $\bm{\mc{C}}(t\leq\tau)$ is used to predict the full correlation function. We note there is some controversy about the precise details of the integral discretization in the context of evaluating path integrals with Trotterized time steps.\cite{Makri2025, Peng2025} Here the kernel is parametrized from the exact data at the larger time step, and no other equations are involved.

\begin{figure}[!b]
    \centering
    \includegraphics[width=\linewidth]{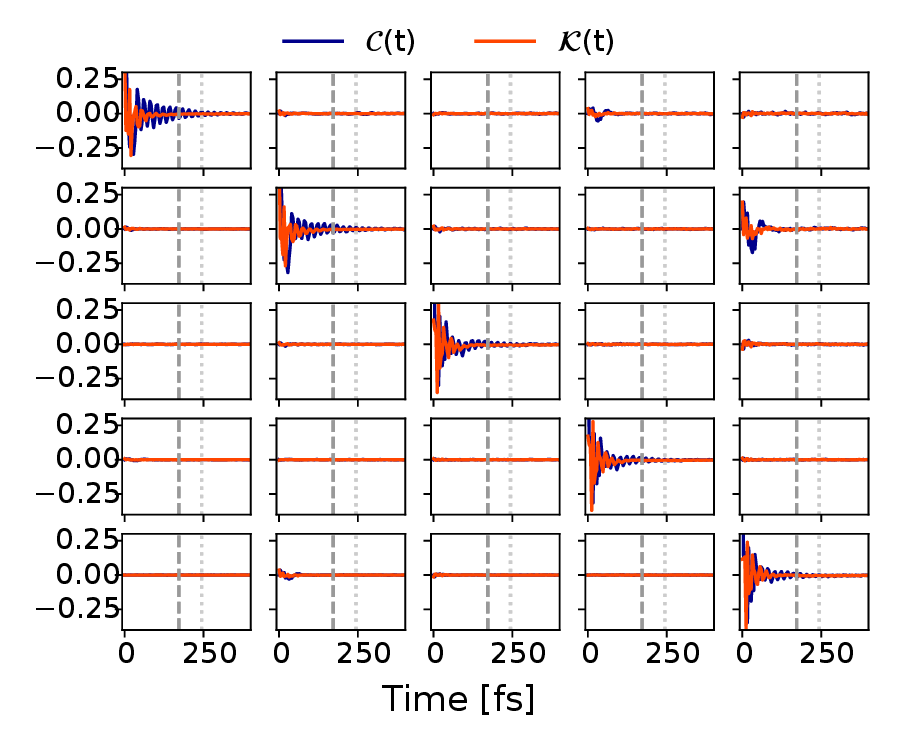}
    \vspace{-18pt}
    \caption{Correlation matrix appropriate to Eq.~\ref{eq:proj_vsfg} and its memory kernel. Dotted line shows when all elements of $\bm{\mc{C}}(t)$ fall below $\epsilon$, which is the first guess at $t_\mathrm{eq}$ (before tapering). Dashed vertical lines shows the kernel cutoff that is subsequently adopted. Time series were normalized before construction, see Appendix~\ref{app:Units}. Elements are arranged in the order $\langle \mathrm{Row}(0) \mathrm{Column}(t)\rangle$ as listed in Eq.~\ref{eq:proj_vsfg}, however multiplication by $(\bm{A}\vert \bm{A})^{-1}$ makes it difficult to precisely identify the panels of the figure with each correlation function, visually. Numerically, the original correlation functions are easily retrieved by undoing the normalization steps.}
    \label{fig:5by5CK}
\end{figure}

\vspace{-4pt}
\section{Results}
\vspace{-8pt}
In generalized Langevin approaches, practitioners often focus on projecting the position and momentum of a particular type of atom, molecules, organism, etc., or some non-linear function of them.\cite{BookBoonYip, BookChapterBerne} The target is something like the crossing of a barrier, the rotation or formation of a bond, a mean distance travelled, and so forth.\cite{Kowalik2019, Ayaz2022, Dalton2025, Klimek2025, KieferArchive} It is equally possible to target a collective property of many molecules,\cite{Lange2006, Hummer2014, Bhattacharyya2024b} as is the case for the susceptibility in Eq.~\ref{eq:corr_2nd_vsfg}. Which coordinates of the full phase space are effectively represented is now unclear and, as such, the minimal amount of information required to obtain $\chi_{ijk}^{(2)}(\omega)$ may not be the choice with the best gain in efficiency. Indeed it is possible that computing additional observables is required, but in what follows we limit ourselves to those directly accessible from the simulations needed to compute VSFG spectra through the standard route,\cite{Morita2006} as this represents a clear-cut case of when the GME method can provide improvements in efficiency. Further work will be needed to determine if the results of parallel simulations could be included; the savings would have to be significant to justify running additional AIMD trajectories.

\begin{figure*}[!t]
    \setcounter{figure}{2}
    \centering
    \includegraphics[width=\linewidth]{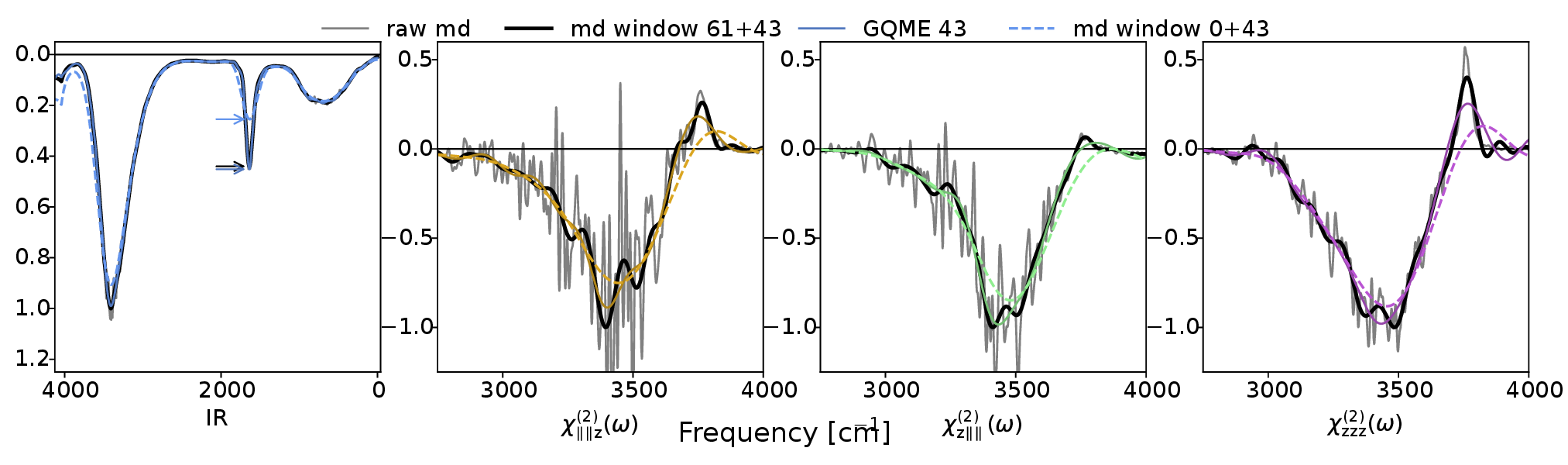}
    \caption{The four predicted spectra using the projector of Eq.~\ref{eq:proj_vsfg} and lengths of correlation function as defined in the legend of 43 (color) and 104~(black) steps respectively. The grey line is with no window applied. The dashed, coloured line is simple windowing while solid coloured line uses the GME. Sampling steps are 4~fs in duration and the underlying timestep was 0.5~fs. In the IR spectrum the arrows mark the peak heights of the central band, with the dashed line underestimating by about half. For the three VSFG spectra, only the stretching region is shown for clarity.}
    \label{fig:5by5_spectra}
\end{figure*} 

\subsection{Minimal VSFG Projector}
\vspace{-8pt}
To begin we define the minimal projection operator needed to compute the 3 spectra $\chi^{(2)}_{\parallel \parallel z}(\omega)$, $\chi^{(2)}_{z \parallel \parallel}(\omega)$, and $\chi^{(2)}_{zzz}(\omega)$, where $x,y \rightarrow \parallel$ is the average of the two in-plane directions which are equivalent by symmetry. Therefore 
\begin{equation}\label{eq:proj_vsfg}
    A \in \lbrace \mu_\parallel, \mu_z, \alpha_{\parallel \parallel}, \alpha_{z \parallel} , \alpha_{z z} \rbrace
\end{equation}
defines a matrix $(\bm{A}(0)|\bm{A}(t))$ where the upper $2\times2$ is block-diagonal with both elements giving the IR spectrum (of the slab), the lower $3\times3$ contains information required for the different polarizations of the Raman spectrum, and the (lower) coupling $3\times 2$ block contains the VSFG spectra. For the numerics, the matrix that is described by the GME in Eq.~\ref{eq:MNZ} is further rotated so that $\bm{\mc{C}}(0)=\mathbb{I}$ with the normalizing factor $(\bm{A}|\bm{A})^{-1}$, which means $(\bm{A}(0)|\bm{A}(0))$ has to be invertible, and indeed we find that it is.

In Figure~\ref{fig:5by5CK} we present $\bm{\mc{C}}(t)$ and its accompanying $\bm{\mc{K}}(t)$ with y-limits chosen to emphasise the behaviour approaching $\tau$ and $t_\mathrm{eq}$. Unlike the earlier cited examples of Ref.~\onlinecite{Lange2006, Carof2014a, Lesnicki2016, Kowalik2019, Vroylandt2022a, Vroylandt2022b, KieferArchive}, in our data we find that the kernel is \textit{not} longer lived than the corresponding $\bm{\mc{C}}(t)$. Indeed, the diagonal elements $\mc{K}_{00},\mc{K}_{11}\rightarrow 0$ much faster than the corresponding elements of the correlation function. However the other elements, particularly the off-diagonal entries, appear to have equivalent lifetimes. Disappointingly then, we might therefore expect no gain in efficiency. To decide if there is any reduction in cost, we compose an error metric. Any metric must ultimately be judged by comparing a `converged' result with spectra predicted using Eq.~\ref{eq:TTM} with varying cutoff $\tau$, but what numerical value constitutes an acceptable threshold is difficult to quantify. The `goodness' of the resulting spectrum is some combination of peak positions, line shape, and band intensity ratios that will depend on exactly for what purpose the scientist wishes to use the result. Nevertheless, we present a quantitative measure of error in Fig.~\ref{fig:5by5_error}.

\begin{figure}[!t]
    \vspace{-8pt}
    \setcounter{figure}{1}
    \centering
    \includegraphics[width=0.9\linewidth]{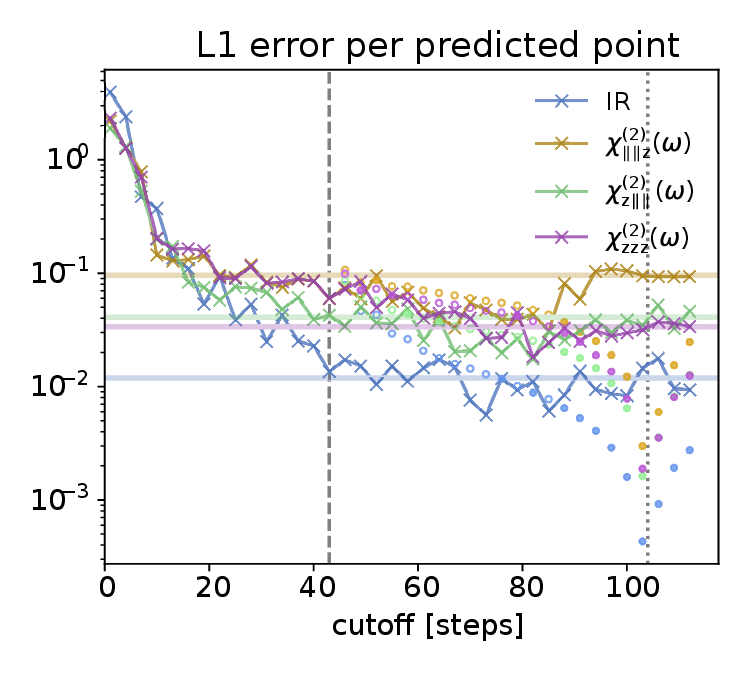}
    \caption{Error defined as the pointwise difference between the `converged' spectrum with window-to-zero ending at $104$~steps (dotted line), and the predicted spectrum with an earlier/later window-to-zero (circles) or produced from GME with cutoff $\tau$ and then windowed (crosses). Circles are filled when they have lower error than the corresponding GME result. The error is normalized by the number of predicted points in time. Horizontal lines represent the average GQME error over the last 5 points, where there is a plateau. Dashed vertical line as in Fig.~\ref{fig:5by5CK}; by error the cutoff is longer for $\chi_{zzz}^{(2)}$, but the other spectra are already converged.}
    \label{fig:5by5_error}
    \vspace{-10pt}
    \setcounter{figure}{3}
\end{figure}
We define our converged result by choosing $t_\mathrm{eq}$ when the correlation functions fall below some sufficiently small value, $|\mc{C}_{ij}(t>t_\mathrm{eq})| < \epsilon~\forall~i,j$ where $\epsilon = 0.015$ by visual inspection, which is 61 steps. We then apply the tapering function for a total length of 104 steps. By comparing deviations the deviation between this converged result with the spectra produced by the GME, Eq.~\ref{eq:TTM}, with varying cutoff, we can inspect the plateau of the curves to work backwards to a reasonable earliest cutoff, which we report to be 43 steps. The GME result at each cutoff can also be compared to the `underconverged' spectrum that would have resulted if the standard procedure using the tapering function had only seen this amount of data. That is, the tapering function reaches zero at $\tau$. We also plot the underconverged deviations in Fig.~\ref{fig:5by5_error}, such that when the GME has a smaller error for a particular cutoff, it represents an improvement in accuracy. As it is not clear exactly when the noise takes over, the original $\epsilon$ is only a guess at the optimal cutoff. From our plot, we know when the `underconverged' error reaches the final GME error, and so we can use that as an improved lower bound on the total data requirement: 75 steps, down from 104. Either way, we find that while the GME method does improve accuracy for correlation functions of duration less than the converged result's, the efficiency savings are minor. We outline how one would realize a cost reduction practically in Appendix~\ref{app:cost_reduction}. 

The spectra to be compared visually are presented in Fig.~\ref{fig:5by5_spectra}. For the IR spectrum, the disagreement mainly stems from the inaccurate band intensity ratio, whilst in the VSFG there is slightly more error in the lineshape of the stretching region. We stop to briefly comment on the $3600$~cm$^{-1}$ shoulder in the $\chi_{\parallel\parallel z}^{(2)}(\omega)$ spectrum, which the GME does not capture but is visible in the tapered version and the experiment. As noted in Sec.~\ref{sec:methods}, the ability to see this feature depends on the choice of tapering function, so we might wonder whether or not it is really presented in our simulation. Absence of this shoulder has been cited\cite{Ohto2015} as an artifact introduced by using the (ss)VVCF\cite{Khatib2016,Khatib2017} approach, which we are not employing. It also does not appear to be a convergence issue, since Ref.~\onlinecite{Ohto2015} could distinguish the feature with only $20$~ps of data while also using fully AIMD $\langle \bm{\mu}(0)\cdot \bm{\alpha}(t)\rangle$ approach. Instead, we attribute differences in our lineshape to temperature. Here, to avoid glassiness in PBE water, we run our simulation at $350$~K, which is higher than Ref.~\onlinecite{Ohto2015} ($320$~K), Ref.~\onlinecite{Khatib2016} ($330$~K), or Ref.~\onlinecite{Litman2023} ($300$~K). Sensitivity of the feature should therefore be expected as it is precisely over this range that the shoulder feature has been shown to disappear when the temperature-dependence was studied using the MB-pol energy function.\cite{Moberg2018}

Overall, the GME performance is quite unexpected, being neither the negative result of a longer-lived memory kernel, nor the massive savings that have been achieved in other contexts. A factor of two reduction in lifetime is pleasant -- we emphasize there is no significant overhead to using the GME-based approach since all elements of the projector are also necessary to compute the spectrum in the established way -- but we can aspire to more. Our objective therefore shifts to trying to find what, if anything, can be done to improve the efficiency gain. In so doing we also wish to understand why it is that a minimal GME fails to achieve significant gains in this system.

\subsection{Augmented and Alternative Projectors}
\vspace{-4pt}
The effect of changing the projector on the resulting GME efficiency is difficult to anticipate when working with collective variables. To interrogate the inclusion or omission of observables we must develop an intuition for how the underlying timescales are feeding through into our projected degrees of freedom.\cite{BookForster} 

To start, we could step back from the VSFG problem and ask a simpler question: what effect does including the polarizibility data $\alpha_{ij}(t)$ have on the efficiency of computing the \textit{infra-red vibrational spectrum}? That is, consider that our $5\times 5$ matrix was constructed by augmenting a smaller matrix containing only $\langle \mu_i(0) \mu_j(t)\rangle$. Whilst the polarizibility may just be the field derivative of the dipole, in recent work on the current-current correlations in polaron forming systems\cite{Bhattacharyya2024b} it was found that a simple\footnote{
Analytically, taking time derivatives of the current begins to include operators describing the phonon bath, and is therefore anything but simple (even unobtainable from the point of view of the HEOM solver). However sufficiently high time resolution allows a numerical estimate.
} 
time derivative of the electric current (collective) autocorrelation was able to reduce cost by up to an order of magnitude. Here, somewhat surprisingly, we find that the prediction of the IR spectrum is insensitive to including the polarizability in the projector, even though the $\langle \alpha_{ij}(0)\mu_k(t) \rangle$ couplings (the VSFG spectra) are non-zero (see Appendix~\ref{app:IRSpectrum}). Thus, it appears that the dipole-dipole correlation functions already contain the information provided by the polarizability time series. The next-slowest motion in the system must be contained in some other observable, but which?

We could look at the problem from the reverse: In the limit that the projector is the identity operator on the full coordinate space, we know a closed system obeys the Schr\"{o}dinger equation, which is Markovian. Our projection is built from operators which act on the phase-space coordinates, discarding some and aggregating others. Each operation that moves our system away from the full phase-space may be responsible for contaminating the memory kernel with slow degrees of freedom, but to our knowledge there is no established convention on how to go about identifying the culprits and so improve the efficiency. One can try to delineate the possibly harmful steps one has unwittingly taken in following a minimal description of the projection operator as follows,
\begin{enumerate}
    \item In computing the dipole moment we include the nuclear positions, but the electronic positions are only included through the first moment of the density. In principle, different electronic densities can have the same dipole moment, so we are not fully describing the electronic position.
    \item Even though we do fully describe the nuclear positions (classically), since we are working with collective quantities in the GME, the cross-terms between coordinates are not included, even though we do calculate them when constructing the total dipole. That is, we are applying the GME only after performing the molecular sum, $\bm{\mu}=\sum_{i=0}^{N-1}\bm{m}_i$.
    \item Our master equation is written purely in terms of time, and not space, as it would be in, say, mode-coupling theory.\cite{BookForster} This happens in the minimal approach because aggregating all molecular contributions into a single value also integrates out position dependence. Yet, we know that the spectral signatures -- the dynamics -- of waters close to the interface differ from those in the bulk. Hence, removing spatial information may have served to increase the memory kernel's lifetime.\cite{KieferArchive}
\end{enumerate}
We will now analyze these steps systematically by quantifying what effect, if any, they have on the efficiency. 

First, we query whether inclusion of higher moments of the density will lead to a more Markovian description. An attractive property of the Voronoi method\cite{Thomas2015} we used to compute the atomic (and then molecular) contributions to the total dipole is that all higher moments of the density are made accessible. This contrasts with the traditional Wannier centre approach,\cite{Gaigeot2003, Sulpizi2012} applied either directly or through machine learning.\cite{Jana2024} We are therefore able to enlarge the projector at no additional cost with the distinct quadrupole moments,
\begin{equation}\label{eq:proj_quad}
    A \in \lbrace \mu_\parallel, \mu_z, \alpha_{\parallel \parallel}, \alpha_{z \parallel} , \alpha_{z z} \rbrace \cup \lbrace Q_{\parallel \parallel}, Q_{z \parallel}, Q_{z z} \rbrace.
\end{equation}
We note that we find $Q_{\parallel z}=Q_{z \parallel}$, which is important because their cross correlations are also identical and so render the time-zero matrix $(\bm{A}\vert\bm{A})$ non-invertible when both are included; there is also a technical point that arises about choice of units, see Appendix~\ref{app:Units}.

\begin{figure}[!b]
    \centering
    \includegraphics[width=\linewidth]{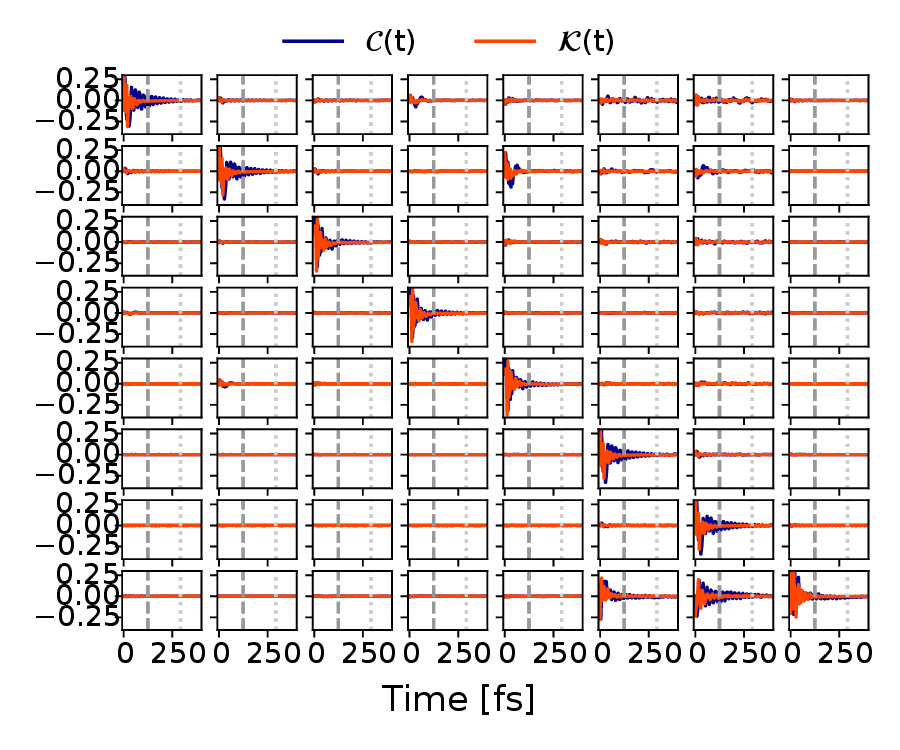}
    \vspace{-24pt}
    \caption{Correlation matrix appropriate to Eq.~\ref{eq:proj_quad}. The upper-left $5 \times 5$ is very close to Fig.~\ref{fig:5by5CK}, but we emphasise they are only equal before (respective) multiplication by the normalizing matrix $(\bm{A}\vert \bm{A})^{-1}$. The lower-right $3 \times 3$ block comes from the new, quadrupole elements. Close inspection reveals a non-zero off-diagonals in the coupling elements to the quadrupole block, however the memory kernel here is very small.}
    \label{fig:8by8CK}
\end{figure}
Figure~\ref{fig:8by8CK} displays the $\bm{\mc{C}}(t)$ and $\bm{\mc{K}}(t)$ matrices for Eq.~\ref{eq:proj_quad}. There is a hint of coupling between the elements linking the first row and the lower $3\times 3$ block ($\mu_\parallel$ and $Q_{ij}$ in the unnormalized matrix respectively) but it is numerically much smaller than the equivalent coupling with the central $3\times 3$ block ($\alpha_{ij}$). Apparently as a result of this, we find predictions of the SFG spectra are unaffected by including quadrupoles in the projector---there are minor quantitative changes to the equivalent of Fig.~\ref{fig:5by5_error}, but no qualitative improvement. Just as including $\alpha_{ij}$ does not improve the efficiency of computing the IR spectrum, including $Q_{ij}$ does not improve the efficiency of computing the VSFG spectra.

\begin{figure}[!ht]
    \vspace{-5pt}
    \centering
    \includegraphics[width=\linewidth]{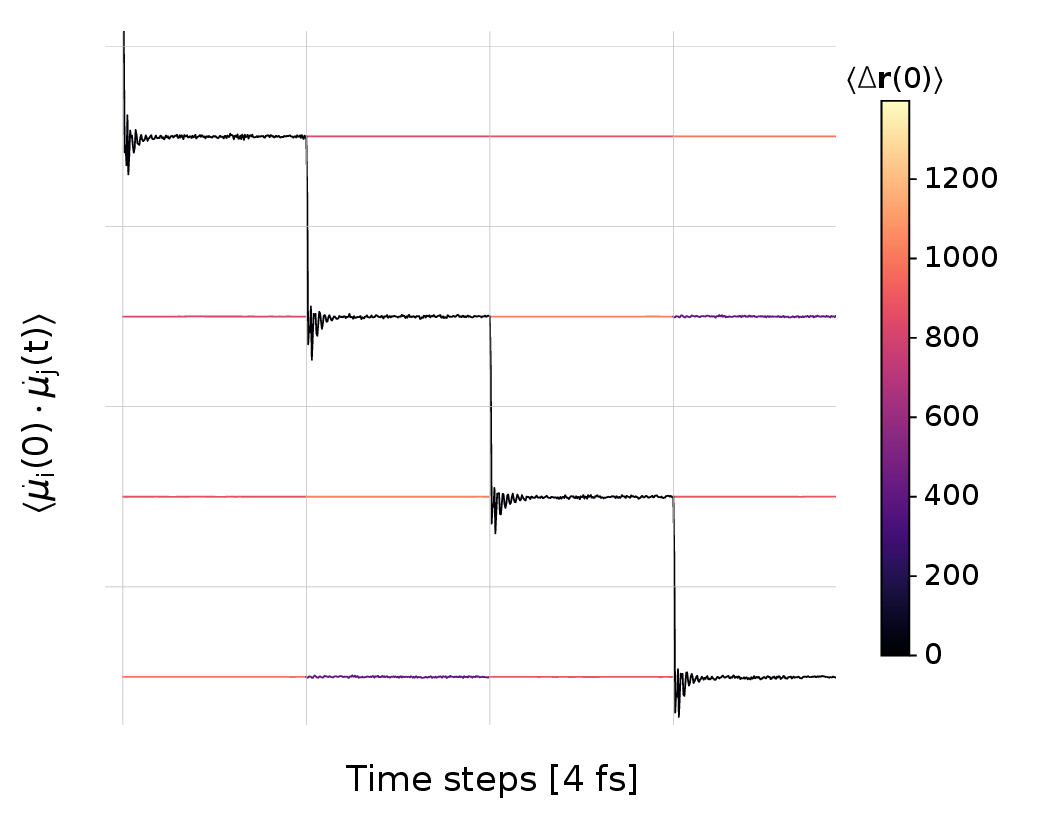}
    \vspace{-5pt}
    \caption{Zoom of a $4 \times 4$ block on the diagonal of the matrix Eq.~\ref{eq:C_molecules}. Color bar shows the average distance between molecules $i$ and $j$ of a particular element over the first frames of the averaging windows used to construct the correlation function (the same value used to determine if a molecule is beyond the cutoff). The full-element sum of this matrix gives the total dipole-dipole correlation function used in previous figures.}
    \vspace{-8pt}
    \label{fig:mu_mu_matrix}
\end{figure}

We therefore move on to our second question: do the set of molecular dipoles $\lbrace\bm{m}_i(t)\rbrace$ serve as a better basis for the projection operator? Certainly, decomposing the collective quantity into molecular level contributions is a chemically intuitive way of viewing the problem and might more directly reveal the nature of the underlying physics. To answer this question, we test whether application of a master equation directly to the molecular correlation matrix, 
\begin{equation}\label{eq:C_molecules}
    C_{ij}(t) = \langle \bm{m}_i(0)\cdot\bm{m}_j(t)\rangle,
\end{equation}
where $i,j$ index the 171~water molecules in the simulation cell, leads to a kernel with a shorter lifetime than starting immediately with the 1-dimensional object $\sum_{\langle i,j\rangle}\langle \bm{m}_i(0)\cdot \bm{m}_j(t)\rangle$.\footnote{
The $\langle i,j\rangle$ notation means when this value is greater than 5~\AA~we zero the element, as described in the Methods section
} 
This matrix can be directly visualized, but at full resolution it is hard to interpret (see Appendix~\ref{app:mu-mu} for reference). We use a colormap to represent the distance between waters $i$ and $j$ at the start of each window, averaged over the whole simulation, and present a zoomed view of a randomly chosen $4\times 4$ block centred on the diagonal in Fig.~\ref{fig:mu_mu_matrix}. Here, we can see that the cross-correlation elements are small compared to the important $<100$~fs, oscillatory region of the self-terms. The rest of the time series we have (previously) discarded as noise. Therefore, the cross correlation elements are of a magnitude and structure very similar to the noisy part of the diagonal elements, even for the waters that are closest in space. It follows that, for these data, operating on the raw matrix is below our fault tolerance. We confirm this by attempting the analysis, and indeed find the GME prediction of the total sum (the full dipole-dipole ACF) is worse than just using the matrix itself. This was to be expected, since it has been shown that these cross-terms require order of nanoseconds to converge.\cite{Ohto2015} So our second question -- whilst motivating advances in simulation technology that unlock cheaper propagation -- must go unanswered with these data. We are left wondering how we can work with noise-containing MD data without completely integrating out the molecular-level information?

The third avenue we identified concerned spatial dependence of the memory. The matrix in Fig.~\ref{fig:mu_mu_matrix} is totally raw and unordered, yet we know that over time there is a similarity between waters at a particular depth due, i.e. isotropy of $x$ and $y$ in a slab system. Indeed, in general, the well-known inhomogeneous broadening of features in the water spectrum arises due to molecules in different local environments. This is particularly important in VSFG since the sign of $\chi^{(2)}_{ijk}(\omega)$ carries information about the orientation of molecules contributing at frequency $\omega$, which is itself reporting on differences in bonding at varying depth with respect to the interface.\cite{Fellows2024} The question of spatial dependence therefore goes hand-in-hand with that of molecular cross-correlation and defining objects that are minimally aggregated whilst still containing acceptably low noise levels.

\begin{figure}[!th]
    \vspace{-8pt}
    \centering
    \includegraphics[width=\linewidth]{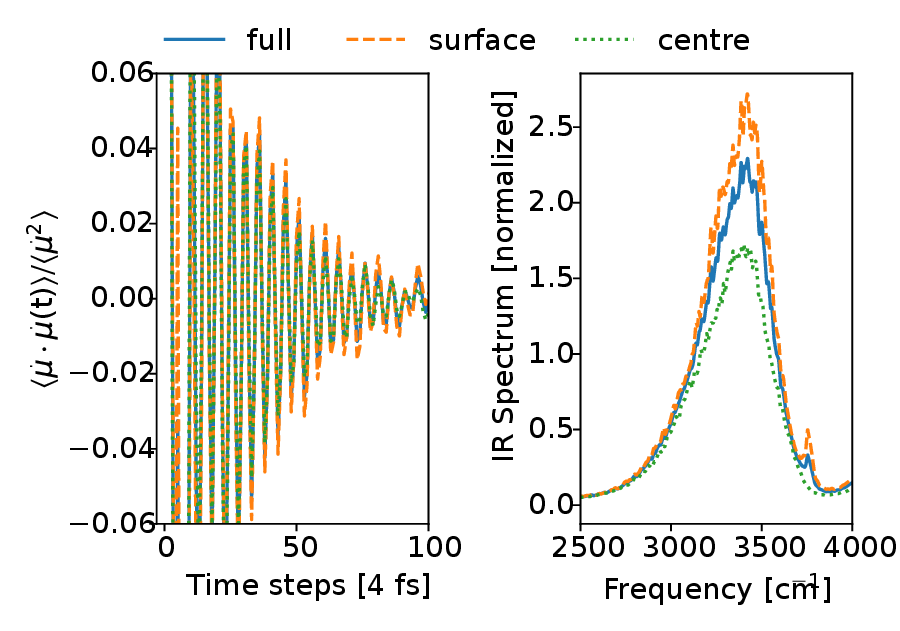}
    \vspace{-8pt}
    \caption{Dipole-dipole correlation function (left) and resulting spectrum after windowing (right) when considering molecules in the whole slab (blue solid), within 5~\AA~of the centre at the start of an averaging window (green dotted), and further than 5~\AA~(orange dashed). Each spectrum is normalized to the feature at 1650~cm$^{-1}$.}
    \label{fig:IRlayers}
\end{figure}

To see if this might be a fruitful line of inquiry we first compute the total dipole autocorrelation (appropriate to the IR spectrum) in two parts: from contributions within 5~\AA~of the centre of the slab, and those further away (within the lower and upper, interfacial layers). In Fig.~\ref{fig:IRlayers} we show side-by-side the time and frequency domain measures for the surface, centre, and combined contributions. As expected, the surface contribution provides an enriched population with fluctuations to the blue, sharply centred around $\sim3800$~cm$^{-1}$, representing not-fully-coordinated O--H stretches.\cite{Auer2008, Stiopkin2011} Yet, whilst they are distinct, the functions in the time-domain decay on what seems to be an identical timescale. Indeed we can compute the associated memory kernels, and find they also have equivalent lifetimes. Now, it may be that only when the coupling between the regions is included that the kernel lifetime is reduced, so we also construct the matrix with elements 
\begin{equation}\label{eq:C_layers}
    C_{uv}=\langle \bm{\mu}_u(0)\cdot\bm{\mu}_v(t)\rangle
\end{equation} 
for $u, v \in \{\mathrm{surface}, \mathrm{centre}\}$. This is defining a GME where the projected observables are total dipole moments arising from particular layers.  We find the off-diagonal elements are of very small magnitude (see Appendix~\ref{app:mu-mu}, Fig.~\ref{fig:surface_centre}). Here, again, the lifetime is unaffected when the kernel is cutoff and the resulting correlation matrix used to construct the IR spectrum (from the sum of diagonal elements). This stands to reason, since we already know that inter-molecular contributions are much smaller than intra-molecular contributions, and most correlations between surface and central regions come from inter-molecular contributions. 

There is a slight nuance to this explanation. Since the correlation function considers the total dipole at time $t$, there is a possibility for molecules to diffuse between the two regions over that time. This leads to intra-molecular contributions to the off-diagonal elements, and these should grow in number as $t$ increases. In spite of this, our result implies that these migrating waters are outnumbered by those that remain in (or leave and re-enter) the region in which they started. Hence, even when the coupling of the two layers is included, there is no reduction in the kernel lifetime. We repeat this approach for other numbers of evenly spaced stratifications, up to 7~layers (see Appendix~\ref{app:mu-mu}, Fig.~\ref{fig:4layers}). As the number of layers increases, the residence probability decreases and the intra-molecular contribution to the off-diagonals grows. Still, we find no improvement in efficiency. If spatial dependence of the memory is able to improve our methodology, it will require a more advanced treatment.\cite{Vroylandt2022b, Kiefer2025, Ayaz2022}

\section{Conclusions}
\vspace{-8pt}
Our first result was a significant one: spectroscopically relevant correlation functions directly calculated from AIMD can benefit from an increase in efficiency using a GME. This is in contrast to many studies where the lifetime of the memory kernel is in excess of that for the correlation function. The GME approach comes at negligible additional cost to running the simulations, even for reduced cost methods, and simple routes to calculating the time non-local kernel -- including the TTM method used in this paper -- are simple to implement with a few lines of code. Yet, the gain in efficiency is not nearly as impressive a result as the orders of magnitude we have come to expect in other systems. 

This challenge of how best to define observables in the projection operator to maximize efficiency is a general one. In biophysics the construction of good collective variables is a central problem\cite{Hummer2014,Dominic2023} and there are many approaches ranging from bespoke problem-specific analysis to data-driven, machine-learning-based methods.\cite{Brotzakis2018,Mardt2018,Chennakesavalu2023,Devergne2025} Indeed, direct prediction of spectra based on structure -- or a set of quantities that are a proxy for the structure -- constitutes an orthogonal approach to spectroscopic computation.\cite{Ye2019,Chen2020,Farahvash2020,Kelly2025} Separately, the problem of how to best construct the memory function in glassy dynamics specifically confronts spatial heterogeneity arising in the liquid state.\cite{Janssen2018} The difficulty of all these interrelated questions advises that future work would benefit from climbing down from the full, electronic-structure problem and working at the FFMD rung of the theoretical hierarchy, first pinning-down treatment of the nuclear contribution to the position correlation function at reduced cost. The trickier electronic description can then be added on subsequently. The need for a principled approach to finding the optimal projector for atomistic simulations of spectroscopic quantities is clear when we consider that expressions for higher-order susceptibilities do not even have straightforward quantum-to-classical replacements, including other SFG spectroscopies that are electronically on-resonance (see Appendix~\ref{app:SFG_derivation}).\cite{Jung2018}

In sum, despite the multitude of different routes we employed to find a better projector, even for just the IR spectrum, the lifetime of the kernel was found to be extremely robust. Whether it be the GME defined on Eq.~\ref{eq:proj_vsfg}, Eq.~\ref{eq:proj_quad}, Eq.~\ref{eq:C_molecules}, or Eq.~\ref{eq:C_layers} we cannot gain a reduction in lifetime larger than $\sim50\%$. This is an interesting result that poses a clear question: what, if any, projector can practically be defined on the output of AIMD simulations to get closer to the Markovian limit of perfectly efficiency spectroscopic prediction?

\vspace{-8pt}
\begin{acknowledgments}
\vspace{-8pt}
T.S.~is the recipient of an Early Career Fellowship from the Leverhulme Trust. Andr\'{e}s Montoya-Castillo, David Wilkins, and Yair Litman are thanked for helpful discussions. This work has made use of the Hamilton HPC Service of Durham University. Funding in the form of an Early Career Fellowship from the Leverhulme Trust (ECF-2024-688) is gratefully acknowledged. 
\end{acknowledgments}

\section*{AUTHOR DECLARATIONS}
\vspace{-12pt}
\section*{Conflict of Interest}
\vspace{-8pt}
The author has no conflicts to disclose.
\vspace{-12pt}

\section*{DATA AVAILABILITY}
\vspace{-8pt}
The data that support the findings of this study are available
from the corresponding author upon reasonable request.

\appendix
\section{Derivation of the SFG correlation function}\label{app:SFG_derivation}
\vspace{-8pt}
The derivation has two steps: first we motivate the correlation of the polarizability with the dipole moment as the correct function to be Fourier transformed, and second we show how it can be replaced with its classical counterpart.

We begin assuming the well-known perturbation theory result for the n$^\mathrm{th}$ order polarization of a quantum density responding to an external field, $\mathrm{Tr}\lbrace \rho^{(n)}\mu \rbrace$. The second order susceptibility, which is the collection of frequency permutations\cite{BookMukamel}
\begin{equation}
    \chi^{(2)}(\omega_1, \omega_2) = \frac{1}{2!}\rho_0\sum_p S^{(2)}(\omega_1+\omega_2, \omega_1)
\end{equation}
of the non-linear response function
\begin{equation}\label{eq:commutators}
    S^{(2)}(t_2, t_1) = \left(\frac{\imi}{\hbar}\right)^2\langle [[V(t_2,t_1), V(t_1)],V(0)]\rho(-\infty)\rangle,
\end{equation}
where $V$ is the light-matter coupling strength which under the usual approximation is the dipole and we understand $t_1$ and $t_2$ are time-ordered (and non-negative). Introducing the quantity $\imi \delta$ to define $I_{vv'}(\omega)\equiv 1/{\omega-\omega_{vv'}+i\delta}$ for transition frequencies $\omega_{vv'}$ (which is taking the matter eigenbasis we cannot actually calculate, and so in practice we take $\delta \rightarrow \Gamma_{vv'}$ as phenomonological damping coefficients), the Fourier-representation can be used to express the response as a sum of Feynman pathways
\begin{equation}
\begin{split}
    \chi^{(2)}(\omega_1, \Omega) = \frac{1}{2}&\left(\frac{1}{\hbar}\right)^2\sum_p\sum_{a,b,c}\rho_0^{(a)}\mu_{ab}\mu_{bc}\mu_{ca} \times \\
    &\big[ I_{ca}(\Omega)I_{ba}(\omega_1) - I_{bc}(\Omega)I_{ba}(\omega_1) \\
    &+ I_{ab}(\Omega)I_{ac}(\omega_1) - I_{bc}(\Omega)I_{ac}(\omega_1) \big]
\end{split}
\end{equation}
where $\Omega=\omega_1+\omega_2$ is the sum frequency. In Hilbert space the two permutations give eight separate terms: half the terms contain the difference between the transition frequency and the radiation, and half the sum. Of these, half contain $\omega_1$ and the other half $\omega_2$.\cite{BookMorita} Since in VSFG only the incoming IR pulse is usually on-resonance with a transition, these two sets are referred to as the resonant and non-resonant contributions. Continuing with only the $\omega_2$ resonant part, 
\begin{equation}
\begin{split}
    \chi^{(2),\mathrm{res}}_{pqr}(\Omega, \omega_2)&=\left(\frac{1}{\hbar}\right)^2\sum_{g,n,m}\rho_0^{(g)}\times\\
    &\Bigg[\frac{\mel{g}{\mu_p}{n}\mel{n}{\mu_q}{m}\mel{m}{\mu_r}{g}}{(\Omega-\omega_{ng}+\imi\Gamma_{ng})(\omega_2-\omega_{mg}+\imi\Gamma_{mg})} \\
    &-\frac{\mel{g}{\mu_q}{m}\mel{m}{\mu_p}{n}\mel{n}{\mu_r}{g}}{(\Omega+\omega_{ng}+\imi\Gamma_{ng})(\omega_2+\omega_{mg}+\imi\Gamma_{mg})}\\
    &+\frac{\mel{g}{\mu_r}{m}\mel{m}{\mu_q}{n}\mel{n}{\mu_p}{g}}{(\Omega-\omega_{nm}+\imi\Gamma_{nm})(\omega_2-\omega_{ng}+\imi\Gamma_{ng})}
    \\
    &-\frac{\mel{g}{\mu_r}{m}\mel{m}{\mu_p}{n}\mel{n}{\mu_q}{g}}{(\Omega-\omega_{nm}+\imi\Gamma_{nm})(\omega_2+\omega_{mg}+\imi\Gamma_{ng})}\Bigg],
\end{split}
\end{equation}
where $g$ and $m$ are in the electronic ground state but can have different vibrational states.
To reveal the Raman tensor, we factorize out $\mel{m}{\mu_r}{g}/(\omega_2-\omega_{mg}+i\Gamma_{mg})$ from all four terms, transposing to
\begin{equation}\label{eq:chi_res_double}
\begin{split}
    \chi^{(2),\mathrm{res}}_{pqr}(\Omega, \omega_2)=-\frac{1}{\hbar}\sum_{g,m}\left(\rho_0^{(g)}-\rho_0^{(m)}  \right)\times
    \\ \frac{\mel{g}{\alpha_{pq}(\Omega)}{m}\mel{m}{\mu_r}{g}}{\omega_2-\omega_{mg}+i\Gamma_{mg}}
\end{split}
\end{equation}
where
\begin{equation}
\begin{split}\label{eq:raman_matrix_element}
    \mel{g}{\alpha_{pq}(\Omega)}{m} &\equiv \\-\frac{1}{\hbar}\sum_n&\left[\frac{\mel{g}{\mu_p}{n}\mel{n}{\mu_q}{m}}{\Omega-\omega_{ng}+\imi\Gamma_{ng}} -\frac{\mel{g}{\mu_q}{n}\mel{n}{\mu_p}{m}}{\Omega+\omega_{nm}+\imi\Gamma_{nm}}\right] 
\end{split}\end{equation}
is the quantity that appears when considering the scattering from state $g$ to $m$, the diagonal elements of which are themselves the first-order susceptibility.\cite{BookMorita, BookMukamel} For the non-resonant part $\omega_2 \rightarrow \omega_1$ and $pqr\rightarrow prq$ on the right-hand side. It is usual to assume the non-resonant part can be modeled as a background.\cite{BookMorita}

At this point, using similar logic of resonance, it is assumed the Raman tensor's frequency dependence can be ignored over the spectral range that $\omega_{ng}\simeq\omega_{nm}$, and it is replaced by a constant value $\alpha(\Omega=0)$ referred to as the polarizability tensor.\cite{BookMorita} In the time domain, completeness over $|m\rangle\langle m|$ can be used to eliminate the second summation in Eq.~\ref{eq:chi_res_double} to give a quantum trace giving the correlation of $\alpha(t)$ with $\mu(0)$.\cite{Auer2008} Indeed, given these approximations on $\Omega$ one can return to Eq.~\ref{eq:commutators} and simplify, instead defining the starting point as\cite{Pouthier1999}
\begin{equation}
    \chi^{(2)}_{pqr}(\omega_1)\simeq-\frac{1}{\hbar}\mathrm{Tr}\lbrace \alpha_{pq} I(\omega_1)[\mu_r, \rho_0] \rbrace,
\end{equation}
where the remaining commutator is moved onto the initial density matrix.

To move between this quantum-mechanical expression for the susceptibility and something appropriate to classical nuclei, we first perform the Kubo transform.\cite{Bader1994} That is, we define the correlation function\cite{BookMahan}
\begin{equation}
    \Phi^{(2)}_{pqr}(t) = \frac{1}{\beta}\int_0^\beta \dd{\beta'} \langle \alpha_{pq}(t-\imi\hbar\beta') \mu_r \rangle
\end{equation}
whose Fourier transform is
\begin{equation}
    \Phi^{(2)}_{pqr}(\omega) = \frac{1-\e{-\beta\hbar\omega}}{\beta\hbar\omega}\int_{-\infty}^\infty\dd{t}\langle\alpha_{pq}(t)\mu_{r}\rangle\e{\imi\omega t}
\end{equation}
which follows from the Fourier multiplier property of the imaginary time translation\cite{BookForster} (shift) justified for the imaginary argument by the KMS condition assuring analyticity between the real axis and $\imi\hbar\beta$ (and further that there is rapid decay at $t\rightarrow\infty$)\cite{BookMahan}. The response function can be written in terms of this `Kubo correlation function' because the commutator with the initial density can be rewritten as a time derivative, since by the chain rule
\begin{equation}
    \frac{\partial}{\partial\lambda}A(\lambda) \equiv \frac{\partial}{\partial\lambda}\left( 
\e{\lambda H} A \e{-\lambda H} \right) = -[H, A(\lambda)],
\end{equation}
and the temperature integral of this is the original commutator (up to a Boltzmann factor).\cite{BookMahan} 
In other words, one can move between 
\begin{equation}
-\frac{\imi}{\hbar}\int_0^\infty\dd{t}\mathrm{Tr}\lbrace\alpha_{pq}(t)[\mu_{r},\rho_0]\rbrace\e{\imi\omega_1 t}
\end{equation}
and
\begin{equation}
    \frac{\imi\omega_1}{2}\left(\Phi^{(2)}_{pqr}(\omega_1) +\frac{\imi}{\pi}P_p\int\dd{\omega}\frac{\Phi^{(2)}_{pqr}(\omega)}{\omega_1-\omega} \right)
\end{equation}
where the principle part is the boundary term that accounts for moving between Laplace and Fourier transforms (required from causality but often omitted since only the imaginary part is desired).\cite{Pouthier1999} The reason this allows correspondence to a classical result is because the Kubo-transformed function has the classical symmetries---the imaginary time translation-plus-integration, which is the `detailed balance prefactor' in frequency space, expresses the quantum correlation function as an anti-commutator using well-known identities, which means it becomes real and even; it reduces to the classical function in the limit $\hbar\rightarrow 0$. 

Other quantum corrections are available, but this is thought to be the best.\cite{Ramirez2004} Note that Ref.~\onlinecite{Litman2023} does perform this replacement, but there is a typo in their equation for the susceptibility.\footnote{Confirmed in a private communication with the authors.} In fact we tested to see what effect omitting the prefactor had and saw little effect on the O--H stretching region, as $\omega$ is already large.

We pause to note that this route assumes the one-time nature of the correlation function at the start. In principle when the full frequency dependence is retained then a two-time Kubo transform would need to be performed.\cite{Reichman2000}  To our knowledge there is no literature on what effect ignoring this fact has on the quality of the correspondence. The two-time version is already much more complicated because no simple (Kramers-Kronig) relationship exists between the correlation function's real and imaginary parts, and so previous work has had to resort to approximation.\cite{Jung2018, Jung2022} There is alternative work based on classical TCFs, but it also assumes particular (harmonic) forms to derive the connection.\cite{DeVane2004}

\section{GQME cost reduction protocol}\label{app:cost_reduction}
The GME lifetime of $\tau<t_\mathrm{eq}$ reduces the cost. It does this in two ways. The first improvement is to do with the ensemble average convergence for a trajectory of fixed length. For a given trajectory of length $N$, for a correlation $\langle X(t)X(t+m)\rangle$ there are $N-m$ samples. Therefore each step in a correlation function away from the zero-delay (the variance) has less precision as a $1/\sqrt{N-m}$ standard error. If the final point $t_\mathrm{eq}$ is replaced by some smaller $\tau$, then the same amount of uncertainty in that final point is achieved for a shorter length trajectory $\aleph=N-(t_\mathrm{eq}-\tau$). This gain comes `for free', and we enjoy this improvement in our purely analytical workflow presented in the main text. Note however that if $N$ is very large compared to $t_\mathrm{eq}$ any gain can be neglected, and indeed here $N$ is 1024 while $t_\mathrm{eq}-\tau\simeq t_\mathrm{eq}/2 \approx 50$, so the precision gain is small.

The second cost reduction is to do with the way samples are generated. In the main text we computed one long equilibrium trajectory. Correlation functions were then obtained by chopping the trajectory into windows, treating each separately, and then averaging the result. This avoids the unfavourable scaling of computing the correlation at longer delays (NlogN for FFT). Some statistics are gained by sliding the window over the trajectory with some step, here half each window (so each data point appears in two windows, except at the first and last half-window). Done this way, only correlations exactly of half-window length make use of the entire statistics: those longer than a half-window do not see all the possible lags of the full trajectory (some correlation between different windows is discarded), and those shorter than a half-window are actually oversampled, with duplication between windows of the same pairs of points in the total trajectory. Oversampling effective downweights some samples (that are not duplicated) and worsens the sampling of these points overall; for the shortest lags $\langle X(t)X(t+\delta t)\rangle$, only the points at the ends of windows are not double-counted, so this error actually peaks at the quarter-window mark. As both $\bm{\mc{C}}(t)$ and $\bm{\mc{K}}(t)$ turned out to be less than half-window length, they are fully in the oversampled regime.

This long-trajectory method is a slow procedure from a computational standpoint because it does not make use of the trivial parallelizability of the equilibrium trajectory. Instead, a short trajectory can be run and then used to seed parallel runs from different starting configurations (which themselves can then seed new runs), which allows a greater amount of HPC resources to be leveraged to perform the ensemble average. This naturally lends itself to the above suggestion of cost reduction because the initial simulation used to produce the first generation of seeds can be used to guess the lifetime of the memory function (in exactly the same way as in our methods). If desired these further trajectories can have their duration adjusted to match (or just-exceed) the memory function's, which drops the sliding-window type approach and further minimizes errors arising from propagating long MD trajectories. On the same metric as before -- how much time is required to achieve the sample number of samples in that last point $t_\mathrm{eq}$ versus $\tau$ -- it is clear that a GME that requires, say, half the length of time to define will accumulate statistics twice as fast, as each trajectory is half the length.

\section{Minimal Projector for the IR Spectrum}\label{app:IRSpectrum}
\begin{figure}[!ht]
    \centering
    \includegraphics[width=0.9\linewidth]{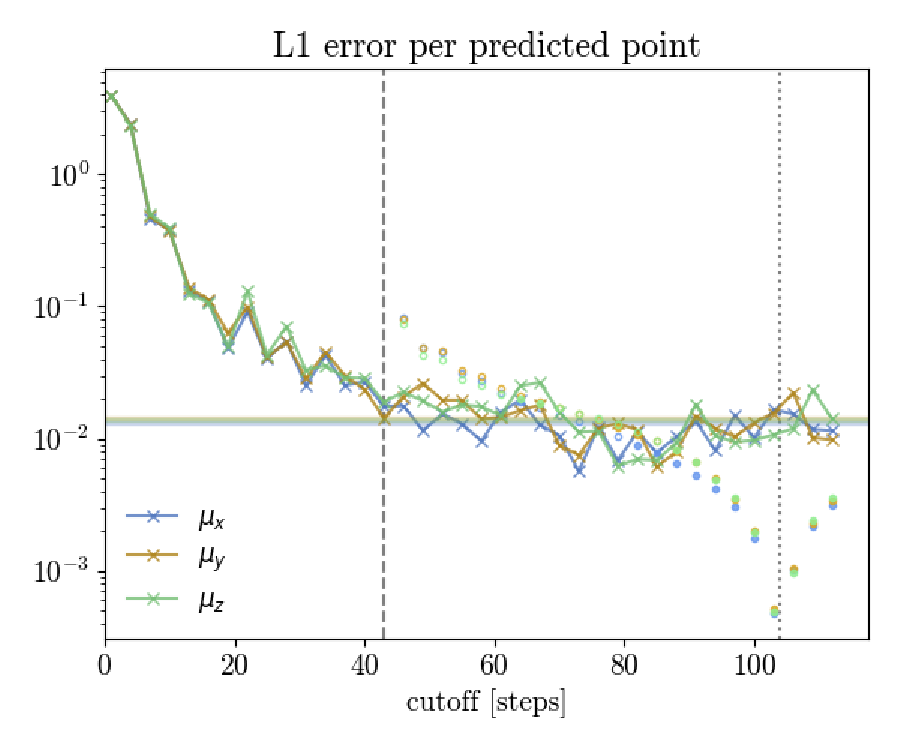}
    \caption{Same as Fig.~\ref{fig:5by5_error} but for only the dipole series. All convergence thresholds are kept the same. Although the precise values at each cutoff are different, the quantitative shape of the error curve is the same in both cases, reaching the same $\sim4\times10^{-2}$ value at long cutoffs.}
    \label{fig:3by3_error}
\end{figure}

Focussing just on the prediction of the IR spectrum from the diagonal $2 \times 2$ block, we ask, is the introduction of the polarization limiting the effectiveness of the GME? This does not seem possible since, as we have described the theory, enlarging the projector at worst keeps the memory kernel lifetime constant. We can test this easily, and indeed from Fig.~\ref{fig:3by3_error} we find that reducing the projector to just the dipole entries (which are completely uncorrelated) yields a very similar error as a function of cutoff as the blue trace in Fig.~\ref{fig:5by5_error}. The lifetime of $K(t)$ given the same convergence criteria stays effective the same, lowering from 35~to 33~steps. This is consistent with our expectation that enlarging the projector cannot decrease the kernel lifetime. The lifetime of $\bm{\mc{C}}(t)$ actually increases from 54~steps to 64~steps.

\section{Combining time series with different units}\label{app:Units}
The factor $(\bm{A}\vert \bm{A})^{-1}$ required for the numerical inversion ensures that the different possible time series of $A$ have appropriate units after projection. One could also normalize time series, effectively changing units, before construction the matrix. Whilst the $\bm{\mc{C}}(t)$ matrix will look different, upon converting back to the original time series (to compute the spectrum) these two approaches should give the same result, and we have confirmed they do for the data of Fig.~\ref{fig:5by5CK}. That is, this does not affect lifetimes or accuracy.

However there can be a numerical difference if the matrix $(\bm{A}\vert \bm{A})$ becomes poorly conditioned. For example $\mu_z$ and $\alpha_{zz}$ may have different orders of magnitude, and the condition number as defined by the ratio of largest to smallest eigenvalues can become very large and render inversion unstable. This is what we found when expanding the projector to include $Q_{ij}$ which are as-computed an additional order of magnitude greater than $\alpha_{zz}$ in these units. For reference, fields were converted to Vpm$^{-1}$ using the factor~2.57. Pre-normalizing the quadrupole time series data removes the issue, and that is what is used in the main text. The normalization factor is stored in case the original units are demanded at a later time.

\section{More correlation matrices}\label{app:mu-mu}
\begin{figure}[!bh]
    \centering
    \includegraphics[width=\linewidth]{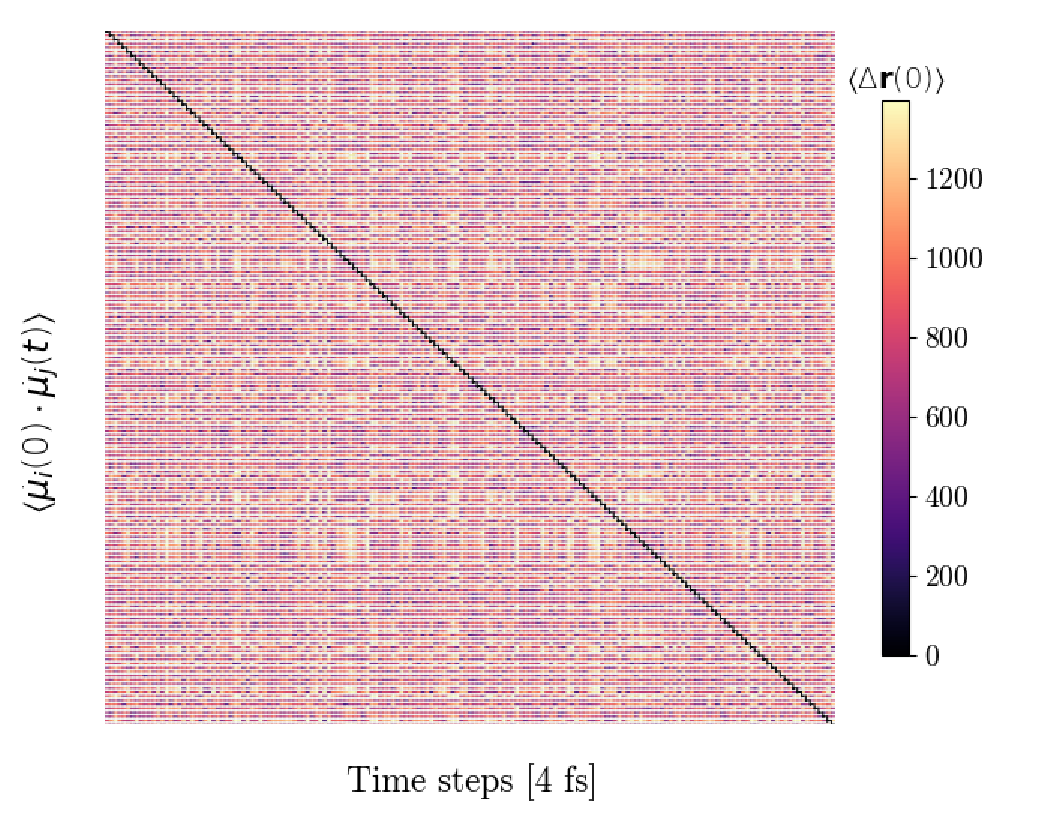}\hspace{-20pt}
    \caption{The matrix Eq.~\ref{eq:C_molecules}, on which a zoom was used to give Fig.~\ref{fig:mu_mu_matrix}.}
    \label{fig:mu_mu_matrix_full}
\end{figure}

In Fig.~\ref{fig:mu_mu_matrix_full} we show the full $\langle \bm{m}_i(0)\cdot \bm{m}_j(t)\rangle $ matrix, which is of dimension $171 \times 171$. The main observations are that 1) the matrix is quite homogeneous, 2) a 5~\AA~cutoff actually includes a large number of the other 170 waters in the box, and 3) most waters that start within 5~\AA~are at a larger distance on average over the length of the full simulation.

\begin{figure}[!h]
    \centering
    \includegraphics[width=0.8\linewidth]{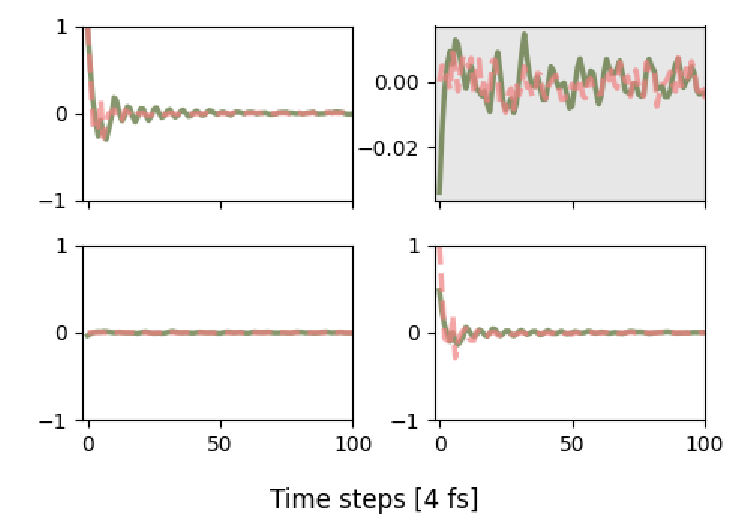}
    \caption{The full surface-centre matrix of Eq.~\ref{eq:C_layers}. Upper-right panel in grey is zoomed version of lower-left panel to show the structure.}
    \label{fig:surface_centre}
\end{figure}

\begin{figure}[!h]
    \centering
    \includegraphics[width=\linewidth]{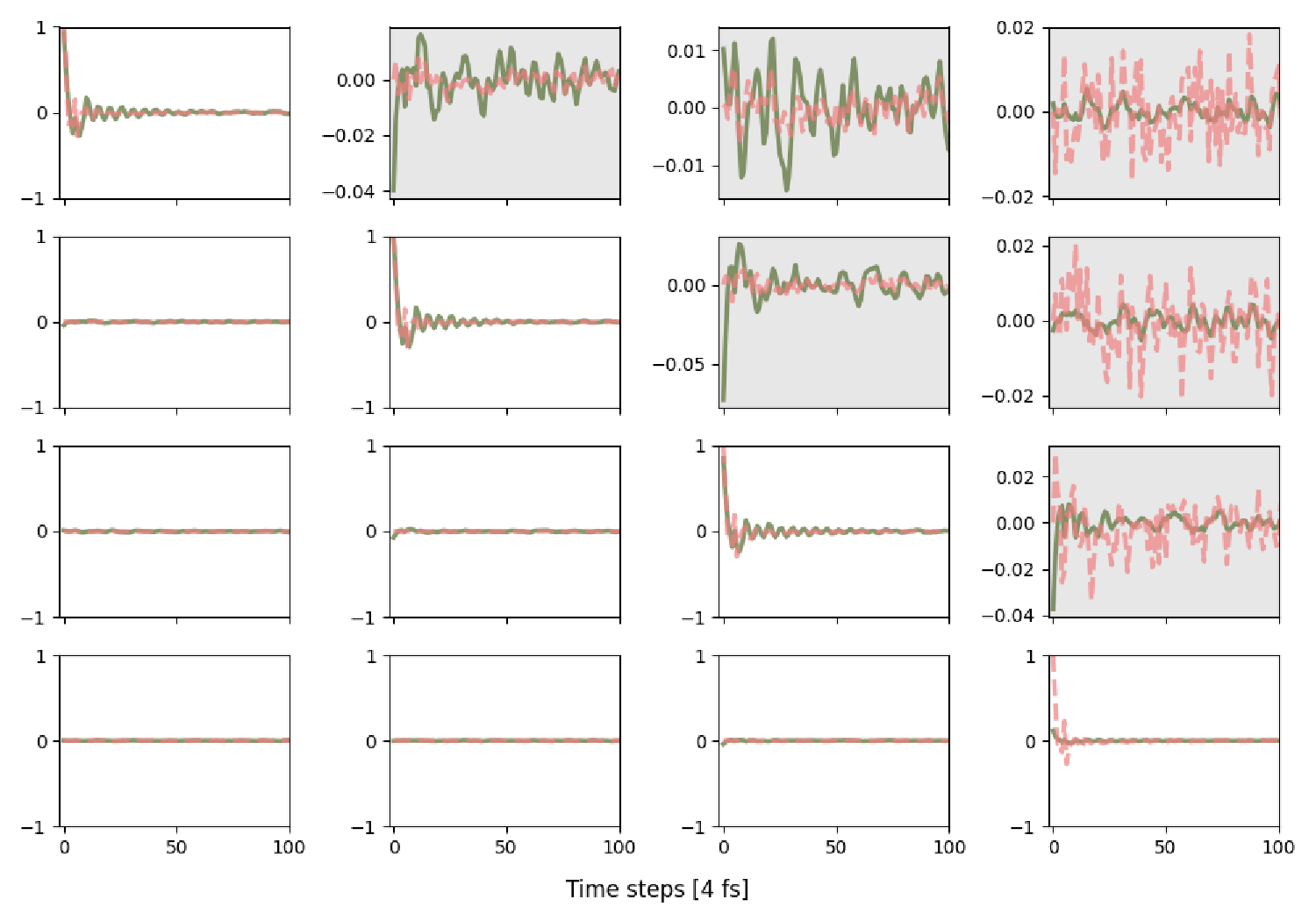}
    \caption{The matrix of Eq.~\ref{eq:C_layers} where $u,v$ denotes molecules in the semi-regular intervals $\lbrace [0.0, 2.5), [2.5, 5.0), [5.0, 7.5) , [7.5,L/2) \rbrace$~\AA~for box length~$L$. Upper-right panels in grey are zoomed versions of lower-left panels to show the structure.}
    \label{fig:4layers}
\end{figure}

In Fig.~\ref{fig:surface_centre} we show the surface-centre coupling projector. As described in the main text, the off-diagonal element is insignificant with its small size. The correlation functions start at different initial values because the numbers of molecules in each region is different. We also constructed similar matrices for larger numbers of layers, and we show the $N=4$ result in Fig.~\ref{fig:4layers}. It is clear that as $N$ increases statistics deteriorate; there are relatively few waters in the top-most layer (bottom right). The layers can be defined to have more equal numbers of molecules but the general conclusions drawn from the diagonal elements -- that the lifetime is the same as the correlation function -- persist.

\pagebreak

\subsection*{References}

\vspace{-22pt}
\bibliography{export}

\begin{thebibliography}{103}%
\makeatletter
\providecommand \@ifxundefined [1]{%
 \@ifx{#1\undefined}
}%
\providecommand \@ifnum [1]{%
 \ifnum #1\expandafter \@firstoftwo
 \else \expandafter \@secondoftwo
 \fi
}%
\providecommand \@ifx [1]{%
 \ifx #1\expandafter \@firstoftwo
 \else \expandafter \@secondoftwo
 \fi
}%
\providecommand \natexlab [1]{#1}%
\providecommand \enquote  [1]{``#1''}%
\providecommand \bibnamefont  [1]{#1}%
\providecommand \bibfnamefont [1]{#1}%
\providecommand \citenamefont [1]{#1}%
\providecommand \href@noop [0]{\@secondoftwo}%
\providecommand \href [0]{\begingroup \@sanitize@url \@href}%
\providecommand \@href[1]{\@@startlink{#1}\@@href}%
\providecommand \@@href[1]{\endgroup#1\@@endlink}%
\providecommand \@sanitize@url [0]{\catcode `\\12\catcode `\$12\catcode `\&12\catcode `\#12\catcode `\^12\catcode `\_12\catcode `\%12\relax}%
\providecommand \@@startlink[1]{}%
\providecommand \@@endlink[0]{}%
\providecommand \url  [0]{\begingroup\@sanitize@url \@url }%
\providecommand \@url [1]{\endgroup\@href {#1}{\urlprefix }}%
\providecommand \urlprefix  [0]{URL }%
\providecommand \Eprint [0]{\href }%
\providecommand \doibase [0]{http://dx.doi.org/}%
\providecommand \selectlanguage [0]{\@gobble}%
\providecommand \bibinfo  [0]{\@secondoftwo}%
\providecommand \bibfield  [0]{\@secondoftwo}%
\providecommand \translation [1]{[#1]}%
\providecommand \BibitemOpen [0]{}%
\providecommand \bibitemStop [0]{}%
\providecommand \bibitemNoStop [0]{.\EOS\space}%
\providecommand \EOS [0]{\spacefactor3000\relax}%
\providecommand \BibitemShut  [1]{\csname bibitem#1\endcsname}%
\let\auto@bib@innerbib\@empty
\bibitem [{\citenamefont {Stevens}\ \emph {et~al.}(2023)\citenamefont {Stevens}, \citenamefont {Grünewald}, \citenamefont {van Tilburg}, \citenamefont {König}, \citenamefont {Gilbert}, \citenamefont {Brier}, \citenamefont {Thornburg}, \citenamefont {Luthey-Schulten},\ and\ \citenamefont {Marrink}}]{Stevens2023}%
  \BibitemOpen
  \bibfield  {author} {\bibinfo {author} {\bibfnamefont {J.~A.}\ \bibnamefont {Stevens}}, \bibinfo {author} {\bibfnamefont {F.}~\bibnamefont {Grünewald}}, \bibinfo {author} {\bibfnamefont {P.~A.}\ \bibnamefont {van Tilburg}}, \bibinfo {author} {\bibfnamefont {M.}~\bibnamefont {König}}, \bibinfo {author} {\bibfnamefont {B.~R.}\ \bibnamefont {Gilbert}}, \bibinfo {author} {\bibfnamefont {T.~A.}\ \bibnamefont {Brier}}, \bibinfo {author} {\bibfnamefont {Z.~R.}\ \bibnamefont {Thornburg}}, \bibinfo {author} {\bibfnamefont {Z.}~\bibnamefont {Luthey-Schulten}}, \ and\ \bibinfo {author} {\bibfnamefont {S.~J.}\ \bibnamefont {Marrink}},\ }\bibfield  {title} {\enquote {\bibinfo {title} {Molecular dynamics simulation of an entire cell},}\ }\href {\doibase 10.3389/fchem.2023.1106495} {\bibfield  {journal} {\bibinfo  {journal} {Frontiers in Chemistry}\ }\textbf {\bibinfo {volume} {11}},\ \bibinfo {pages} {1106495} (\bibinfo {year} {2023})}\BibitemShut {NoStop}%
\bibitem [{\citenamefont {Gaudy}, \citenamefont {Salanne},\ and\ \citenamefont {Merlet}(2024)}]{Gaudy2024}%
  \BibitemOpen
  \bibfield  {author} {\bibinfo {author} {\bibfnamefont {N.}~\bibnamefont {Gaudy}}, \bibinfo {author} {\bibfnamefont {M.}~\bibnamefont {Salanne}}, \ and\ \bibinfo {author} {\bibfnamefont {C.}~\bibnamefont {Merlet}},\ }\bibfield  {title} {\enquote {\bibinfo {title} {Dynamics and energetics of ion adsorption at the interface between a pure ionic liquid and carbon electrodes},}\ }\href {\doibase 10.1021/ACS.JPCB.4C01192} {\bibfield  {journal} {\bibinfo  {journal} {The Journal of Physical Chemistry B}\ }\textbf {\bibinfo {volume} {128}},\ \bibinfo {pages} {5064--5071} (\bibinfo {year} {2024})}\BibitemShut {NoStop}%
\bibitem [{\citenamefont {Frenkel}\ and\ \citenamefont {Smit}(2001)}]{BookFrenkelSmit}%
  \BibitemOpen
  \bibfield  {author} {\bibinfo {author} {\bibfnamefont {D.}~\bibnamefont {Frenkel}}\ and\ \bibinfo {author} {\bibfnamefont {B.}~\bibnamefont {Smit}},\ }\href {https://books.google.co.uk/books?id=5qTzldS9ROIC} {\emph {\bibinfo {title} {Understanding Molecular Simulation: From Algorithms to Applications}}}\ (\bibinfo  {publisher} {Elsevier Science},\ \bibinfo {year} {2001})\BibitemShut {NoStop}%
\bibitem [{\citenamefont {Liberatore}, \citenamefont {Meli},\ and\ \citenamefont {Rothlisberger}(2018)}]{Liberatore2018}%
  \BibitemOpen
  \bibfield  {author} {\bibinfo {author} {\bibfnamefont {E.}~\bibnamefont {Liberatore}}, \bibinfo {author} {\bibfnamefont {R.}~\bibnamefont {Meli}}, \ and\ \bibinfo {author} {\bibfnamefont {U.}~\bibnamefont {Rothlisberger}},\ }\bibfield  {title} {\enquote {\bibinfo {title} {A versatile multiple time step scheme for efficient ab initio molecular dynamics simulations},}\ }\href {\doibase 10.1021/ACS.JCTC.7B01189} {\bibfield  {journal} {\bibinfo  {journal} {Journal of Chemical Theory and Computation}\ }\textbf {\bibinfo {volume} {14}},\ \bibinfo {pages} {2834--2842} (\bibinfo {year} {2018})}\BibitemShut {NoStop}%
\bibitem [{\citenamefont {Andermatt}\ \emph {et~al.}(2016)\citenamefont {Andermatt}, \citenamefont {Cha}, \citenamefont {Schiffmann},\ and\ \citenamefont {VandeVondele}}]{Andermatt2016}%
  \BibitemOpen
  \bibfield  {author} {\bibinfo {author} {\bibfnamefont {S.}~\bibnamefont {Andermatt}}, \bibinfo {author} {\bibfnamefont {J.}~\bibnamefont {Cha}}, \bibinfo {author} {\bibfnamefont {F.}~\bibnamefont {Schiffmann}}, \ and\ \bibinfo {author} {\bibfnamefont {J.}~\bibnamefont {VandeVondele}},\ }\bibfield  {title} {\enquote {\bibinfo {title} {Combining linear-scaling dft with subsystem dft in born–oppenheimer and ehrenfest molecular dynamics simulations: From molecules to a virus in solution},}\ }\href {\doibase 10.1021/ACS.JCTC.6B00398} {\bibfield  {journal} {\bibinfo  {journal} {Journal of Chemical Theory and Computation}\ }\textbf {\bibinfo {volume} {12}},\ \bibinfo {pages} {3214--3227} (\bibinfo {year} {2016})}\BibitemShut {NoStop}%
\bibitem [{\citenamefont {Lawrence}\ \emph {et~al.}(2023)\citenamefont {Lawrence}, \citenamefont {Lieberherr}, \citenamefont {Fletcher},\ and\ \citenamefont {Manolopoulos}}]{Lawrence2023}%
  \BibitemOpen
  \bibfield  {author} {\bibinfo {author} {\bibfnamefont {J.~E.}\ \bibnamefont {Lawrence}}, \bibinfo {author} {\bibfnamefont {A.~Z.}\ \bibnamefont {Lieberherr}}, \bibinfo {author} {\bibfnamefont {T.}~\bibnamefont {Fletcher}}, \ and\ \bibinfo {author} {\bibfnamefont {D.~E.}\ \bibnamefont {Manolopoulos}},\ }\bibfield  {title} {\enquote {\bibinfo {title} {Fast quasi-centroid molecular dynamics for water and ice},}\ }\href {\doibase 10.1021/acs.jpcb.3c05028} {\bibfield  {journal} {\bibinfo  {journal} {J. Phys. Chem. B}\ }\textbf {\bibinfo {volume} {127}},\ \bibinfo {pages} {9180} (\bibinfo {year} {2023})}\BibitemShut {NoStop}%
\bibitem [{\citenamefont {Ginsberg}\ and\ \citenamefont {Tisdale}(2020)}]{Ginsberg2020}%
  \BibitemOpen
  \bibfield  {author} {\bibinfo {author} {\bibfnamefont {N.~S.}\ \bibnamefont {Ginsberg}}\ and\ \bibinfo {author} {\bibfnamefont {W.~A.}\ \bibnamefont {Tisdale}},\ }\bibfield  {title} {\enquote {\bibinfo {title} {Spatially resolved photogenerated exciton and charge transport in emerging semiconductors},}\ }\href {\doibase 10.1146/ANNUREV-PHYSCHEM-052516-050703} {\bibfield  {journal} {\bibinfo  {journal} {Annual Review of Physical Chemistry}\ }\textbf {\bibinfo {volume} {71}},\ \bibinfo {pages} {1--30} (\bibinfo {year} {2020})}\BibitemShut {NoStop}%
\bibitem [{\citenamefont {Giannini}\ \emph {et~al.}(2022)\citenamefont {Giannini}, \citenamefont {Peng}, \citenamefont {Cupellini}, \citenamefont {Padula}, \citenamefont {Carof},\ and\ \citenamefont {Blumberger}}]{Giannini2022}%
  \BibitemOpen
  \bibfield  {author} {\bibinfo {author} {\bibfnamefont {S.}~\bibnamefont {Giannini}}, \bibinfo {author} {\bibfnamefont {W.~T.}\ \bibnamefont {Peng}}, \bibinfo {author} {\bibfnamefont {L.}~\bibnamefont {Cupellini}}, \bibinfo {author} {\bibfnamefont {D.}~\bibnamefont {Padula}}, \bibinfo {author} {\bibfnamefont {A.}~\bibnamefont {Carof}}, \ and\ \bibinfo {author} {\bibfnamefont {J.}~\bibnamefont {Blumberger}},\ }\bibfield  {title} {\enquote {\bibinfo {title} {Exciton transport in molecular organic semiconductors boosted by transient quantum delocalization},}\ }\href {\doibase 10.1038/S41467-022-30308-5} {\bibfield  {journal} {\bibinfo  {journal} {Nature Communications}\ }\textbf {\bibinfo {volume} {13}},\ \bibinfo {pages} {1--13} (\bibinfo {year} {2022})}\BibitemShut {NoStop}%
\bibitem [{\citenamefont {Burke}, \citenamefont {Landi},\ and\ \citenamefont {Troisi}(2024)}]{Burke2024}%
  \BibitemOpen
  \bibfield  {author} {\bibinfo {author} {\bibfnamefont {C.}~\bibnamefont {Burke}}, \bibinfo {author} {\bibfnamefont {A.}~\bibnamefont {Landi}}, \ and\ \bibinfo {author} {\bibfnamefont {A.}~\bibnamefont {Troisi}},\ }\bibfield  {title} {\enquote {\bibinfo {title} {The dynamic nature of electrostatic disorder in organic mixed ionic and electronic conductors},}\ }\href {\doibase 10.1039/D4MH00706A} {\bibfield  {journal} {\bibinfo  {journal} {Materials Horizons}\ }\textbf {\bibinfo {volume} {11}},\ \bibinfo {pages} {5313--5319} (\bibinfo {year} {2024})}\BibitemShut {NoStop}%
\bibitem [{\citenamefont {Bhattacharyya}, \citenamefont {Sayer},\ and\ \citenamefont {Montoya-Castillo}(2024{\natexlab{a}})}]{Bhattacharyya2024a}%
  \BibitemOpen
  \bibfield  {author} {\bibinfo {author} {\bibfnamefont {S.}~\bibnamefont {Bhattacharyya}}, \bibinfo {author} {\bibfnamefont {T.}~\bibnamefont {Sayer}}, \ and\ \bibinfo {author} {\bibfnamefont {A.}~\bibnamefont {Montoya-Castillo}},\ }\bibfield  {title} {\enquote {\bibinfo {title} {Anomalous transport of small polarons arises from transient lattice relaxation or immovable boundaries},}\ }\href {\doibase 10.1021/ACS.JPCLETT.3C03380} {\bibfield  {journal} {\bibinfo  {journal} {The Journal of Physical Chemistry Letters}\ }\textbf {\bibinfo {volume} {15}},\ \bibinfo {pages} {1382--1389} (\bibinfo {year} {2024}{\natexlab{a}})}\BibitemShut {NoStop}%
\bibitem [{\citenamefont {Bhattacharyya}, \citenamefont {Sayer},\ and\ \citenamefont {Montoya-Castillo}(2024{\natexlab{b}})}]{Bhattacharyya2024b}%
  \BibitemOpen
  \bibfield  {author} {\bibinfo {author} {\bibfnamefont {S.}~\bibnamefont {Bhattacharyya}}, \bibinfo {author} {\bibfnamefont {T.}~\bibnamefont {Sayer}}, \ and\ \bibinfo {author} {\bibfnamefont {A.}~\bibnamefont {Montoya-Castillo}},\ }\bibfield  {title} {\enquote {\bibinfo {title} {Mori generalized master equations offer an efficient route to predict and interpret polaron transport},}\ }\href {\doibase 10.1039/D4SC03144J} {\bibfield  {journal} {\bibinfo  {journal} {Chemical Science}\ }\textbf {\bibinfo {volume} {15}},\ \bibinfo {pages} {16715--16723} (\bibinfo {year} {2024}{\natexlab{b}})}\BibitemShut {NoStop}%
\bibitem [{\citenamefont {Libbi}\ \emph {et~al.}(2025)\citenamefont {Libbi}, \citenamefont {Johansson}, \citenamefont {Kozinsky},\ and\ \citenamefont {Monacelli}}]{Libbi2025}%
  \BibitemOpen
  \bibfield  {author} {\bibinfo {author} {\bibfnamefont {F.}~\bibnamefont {Libbi}}, \bibinfo {author} {\bibfnamefont {A.}~\bibnamefont {Johansson}}, \bibinfo {author} {\bibfnamefont {B.}~\bibnamefont {Kozinsky}}, \ and\ \bibinfo {author} {\bibfnamefont {L.}~\bibnamefont {Monacelli}},\ }\bibfield  {title} {\enquote {\bibinfo {title} {Nonequilibrium quantum dynamics in srtio3 under impulsive thz radiation with machine learning},}\ }\href {\doibase 10.1126/SCIADV.ADW1634} {\bibfield  {journal} {\bibinfo  {journal} {Science Advances}\ }\textbf {\bibinfo {volume} {11}},\ \bibinfo {pages} {1634} (\bibinfo {year} {2025})}\BibitemShut {NoStop}%
\bibitem [{\citenamefont {Bhattacharyya}, \citenamefont {Sayer},\ and\ \citenamefont {Montoya-Castillo}(2025)}]{Bhattacharyya2025a}%
  \BibitemOpen
  \bibfield  {author} {\bibinfo {author} {\bibfnamefont {S.}~\bibnamefont {Bhattacharyya}}, \bibinfo {author} {\bibfnamefont {T.}~\bibnamefont {Sayer}}, \ and\ \bibinfo {author} {\bibfnamefont {A.}~\bibnamefont {Montoya-Castillo}},\ }\bibfield  {title} {\enquote {\bibinfo {title} {Space-local memory in generalized master equations: Reaching the thermodynamic limit for the cost of a small lattice simulation},}\ }\href {\doibase 10.1063/5.0249145} {\bibfield  {journal} {\bibinfo  {journal} {The Journal of Chemical Physics}\ }\textbf {\bibinfo {volume} {162}},\ \bibinfo {pages} {91102} (\bibinfo {year} {2025})}\BibitemShut {NoStop}%
\bibitem [{\citenamefont {Berne}(1977)}]{BookChapterBerne}%
  \BibitemOpen
  \bibfield  {author} {\bibinfo {author} {\bibfnamefont {B.~J.}\ \bibnamefont {Berne}},\ }\enquote {\bibinfo {title} {Projection operator techniques in the theory of fluctuations},}\ in\ \href {https://link.springer.com/book/9781461579083} {\emph {\bibinfo {booktitle} {Statistical Mechanics: Part B: Time-Dependent Processes}}},\ Vol.~\bibinfo {volume} {6},\ \bibinfo {editor} {edited by\ \bibinfo {editor} {\bibfnamefont {W.~H.}\ \bibnamefont {Miller}}, \bibinfo {editor} {\bibfnamefont {H.~F.~I.}\ \bibnamefont {Schaefer}}, \bibinfo {editor} {\bibfnamefont {B.~J.}\ \bibnamefont {Berne}}, \ and\ \bibinfo {editor} {\bibfnamefont {G.~A.}\ \bibnamefont {Segal}}}\ (\bibinfo  {publisher} {Plenum Press},\ \bibinfo {year} {1977})\ pp.\ \bibinfo {pages} {233--257}\BibitemShut {NoStop}%
\bibitem [{Note1()}]{Note1}%
  \BibitemOpen
  \bibinfo {note} {Strictly speaking, the dipole moment in a periodic system is not an observable.\cite {Resta1998} Only changes in the total polarization are well-defined. Nevertheless the operator we associate with the dipole (molecular or total) can still be written and computed within the limits of the Berry-phase polarization branch,\cite {Gaigeot2003} provided the system remains insulating\cite {Spaldin2012}}\BibitemShut {NoStop}%
\bibitem [{\citenamefont {Nakajima}(1958)}]{Nakajima1958a}%
  \BibitemOpen
  \bibfield  {author} {\bibinfo {author} {\bibfnamefont {S.}~\bibnamefont {Nakajima}},\ }\bibfield  {title} {\enquote {\bibinfo {title} {On quantum theory of transport phenomena: Steady diffusion},}\ }\href {\doibase 10.1143/PTP.20.948} {\bibfield  {journal} {\bibinfo  {journal} {Progress of Theoretical Physics}\ }\textbf {\bibinfo {volume} {20}},\ \bibinfo {pages} {948--959} (\bibinfo {year} {1958})}\BibitemShut {NoStop}%
\bibitem [{\citenamefont {Zwanzig}(1960)}]{Zwanzig1960}%
  \BibitemOpen
  \bibfield  {author} {\bibinfo {author} {\bibfnamefont {R.}~\bibnamefont {Zwanzig}},\ }\bibfield  {title} {\enquote {\bibinfo {title} {Ensemble method in the theory of irreversibility},}\ }\href {\doibase 10.1063/1.1731409} {\bibfield  {journal} {\bibinfo  {journal} {The Journal of Chemical Physics}\ }\textbf {\bibinfo {volume} {33}},\ \bibinfo {pages} {1338--1341} (\bibinfo {year} {1960})}\BibitemShut {NoStop}%
\bibitem [{\citenamefont {Mori}(1965)}]{Mori1965b}%
  \BibitemOpen
  \bibfield  {author} {\bibinfo {author} {\bibfnamefont {H.}~\bibnamefont {Mori}},\ }\bibfield  {title} {\enquote {\bibinfo {title} {Transport, collective motion, and brownian motion},}\ }\href {\doibase 10.1143/PTP.33.423} {\bibfield  {journal} {\bibinfo  {journal} {Progress of Theoretical Physics}\ }\textbf {\bibinfo {volume} {33}},\ \bibinfo {pages} {423--455} (\bibinfo {year} {1965})}\BibitemShut {NoStop}%
\bibitem [{\citenamefont {Boon}\ and\ \citenamefont {Yip}(1991)}]{BookBoonYip}%
  \BibitemOpen
  \bibfield  {author} {\bibinfo {author} {\bibfnamefont {J.-P.}\ \bibnamefont {Boon}}\ and\ \bibinfo {author} {\bibfnamefont {S.}~\bibnamefont {Yip}},\ }\href@noop {} {\emph {\bibinfo {title} {Molecular Hydrodynamics}}}\ (\bibinfo  {publisher} {Dover Publications},\ \bibinfo {year} {1991})\ p.\ \bibinfo {pages} {417}\BibitemShut {NoStop}%
\bibitem [{\citenamefont {Husic}\ and\ \citenamefont {Pande}(2018)}]{Husic2018}%
  \BibitemOpen
  \bibfield  {author} {\bibinfo {author} {\bibfnamefont {B.~E.}\ \bibnamefont {Husic}}\ and\ \bibinfo {author} {\bibfnamefont {V.~S.}\ \bibnamefont {Pande}},\ }\bibfield  {title} {\enquote {\bibinfo {title} {Markov state models: From an art to a science},}\ }\href {\doibase 10.1021/JACS.7B12191} {\bibfield  {journal} {\bibinfo  {journal} {Journal of the American Chemical Society}\ }\textbf {\bibinfo {volume} {140}},\ \bibinfo {pages} {2386--2396} (\bibinfo {year} {2018})}\BibitemShut {NoStop}%
\bibitem [{\citenamefont {Pfalzgraff}\ \emph {et~al.}(2019)\citenamefont {Pfalzgraff}, \citenamefont {Montoya-Castillo}, \citenamefont {Kelly},\ and\ \citenamefont {Markland}}]{Pfalzgraff2019a}%
  \BibitemOpen
  \bibfield  {author} {\bibinfo {author} {\bibfnamefont {W.~C.}\ \bibnamefont {Pfalzgraff}}, \bibinfo {author} {\bibfnamefont {A.}~\bibnamefont {Montoya-Castillo}}, \bibinfo {author} {\bibfnamefont {A.}~\bibnamefont {Kelly}}, \ and\ \bibinfo {author} {\bibfnamefont {T.~E.}\ \bibnamefont {Markland}},\ }\bibfield  {title} {\enquote {\bibinfo {title} {Efficient construction of generalized master equation memory kernels for multi-state systems from nonadiabatic quantum-classical dynamics},}\ }\href {\doibase 10.1063/1.5095715} {\bibfield  {journal} {\bibinfo  {journal} {The Journal of Chemical Physics}\ }\textbf {\bibinfo {volume} {150}},\ \bibinfo {pages} {244109} (\bibinfo {year} {2019})}\BibitemShut {NoStop}%
\bibitem [{\citenamefont {Sayer}\ and\ \citenamefont {Montoya-Castillo}(2024)}]{Sayer2024}%
  \BibitemOpen
  \bibfield  {author} {\bibinfo {author} {\bibfnamefont {T.}~\bibnamefont {Sayer}}\ and\ \bibinfo {author} {\bibfnamefont {A.}~\bibnamefont {Montoya-Castillo}},\ }\bibfield  {title} {\enquote {\bibinfo {title} {Efficient formulation of multitime generalized quantum master equations: Taming the cost of simulating 2d spectra},}\ }\href {\doibase 10.1063/5.0185578} {\bibfield  {journal} {\bibinfo  {journal} {The Journal of Chemical Physics}\ }\textbf {\bibinfo {volume} {160}},\ \bibinfo {pages} {44108} (\bibinfo {year} {2024})}\BibitemShut {NoStop}%
\bibitem [{\citenamefont {Trenins}\ and\ \citenamefont {Rossi}(2025)}]{Trenins2025}%
  \BibitemOpen
  \bibfield  {author} {\bibinfo {author} {\bibfnamefont {G.}~\bibnamefont {Trenins}}\ and\ \bibinfo {author} {\bibfnamefont {M.}~\bibnamefont {Rossi}},\ }\bibfield  {title} {\enquote {\bibinfo {title} {Non-markovian effects in quantum rate calculations of hydrogen diffusion with electronic friction},}\ }\href {\doibase 10.1103/PhysRevLett.134.226201} {\bibfield  {journal} {\bibinfo  {journal} {Physical Review Letters}\ }\textbf {\bibinfo {volume} {134}},\ \bibinfo {pages} {226201} (\bibinfo {year} {2025})}\BibitemShut {NoStop}%
\bibitem [{Note2()}]{Note2}%
  \BibitemOpen
  \bibinfo {note} {The authors of Ref.~\protect \rev@citealpnum {Trenins2025} have informed us the lifetime of the RPMD kernel is just 25~fs, two orders of magnitude shorter than the transmission coefficient plateau time in the low-friction regime.}\BibitemShut {Stop}%
\bibitem [{\citenamefont {Cao}\ \emph {et~al.}(2020)\citenamefont {Cao}, \citenamefont {Montoya-Castillo}, \citenamefont {Wang}, \citenamefont {Markland},\ and\ \citenamefont {Huang}}]{Cao2020a}%
  \BibitemOpen
  \bibfield  {author} {\bibinfo {author} {\bibfnamefont {S.}~\bibnamefont {Cao}}, \bibinfo {author} {\bibfnamefont {A.}~\bibnamefont {Montoya-Castillo}}, \bibinfo {author} {\bibfnamefont {W.}~\bibnamefont {Wang}}, \bibinfo {author} {\bibfnamefont {T.~E.}\ \bibnamefont {Markland}}, \ and\ \bibinfo {author} {\bibfnamefont {X.}~\bibnamefont {Huang}},\ }\bibfield  {title} {\enquote {\bibinfo {title} {On the advantages of exploiting memory in markov state models for biomolecular dynamics},}\ }\href {\doibase 10.1063/5.0010787} {\bibfield  {journal} {\bibinfo  {journal} {The Journal of Chemical Physics}\ }\textbf {\bibinfo {volume} {153}},\ \bibinfo {pages} {14105} (\bibinfo {year} {2020})}\BibitemShut {NoStop}%
\bibitem [{\citenamefont {Dominic}\ \emph {et~al.}(2023{\natexlab{a}})\citenamefont {Dominic}, \citenamefont {Sayer}, \citenamefont {Cao}, \citenamefont {Markland}, \citenamefont {Huang},\ and\ \citenamefont {Montoya-castillo}}]{Dominic2022}%
  \BibitemOpen
  \bibfield  {author} {\bibinfo {author} {\bibfnamefont {A.~J.}\ \bibnamefont {Dominic}}, \bibinfo {author} {\bibfnamefont {T.}~\bibnamefont {Sayer}}, \bibinfo {author} {\bibfnamefont {S.}~\bibnamefont {Cao}}, \bibinfo {author} {\bibfnamefont {T.~E.}\ \bibnamefont {Markland}}, \bibinfo {author} {\bibfnamefont {X.}~\bibnamefont {Huang}}, \ and\ \bibinfo {author} {\bibfnamefont {A.}~\bibnamefont {Montoya-castillo}},\ }\bibfield  {title} {\enquote {\bibinfo {title} {Building insightful, memory-enriched models to capture long-time biochemical processes from short-time simulations},}\ }\href {\doibase 10.1073/pnas.2221048120} {\bibfield  {journal} {\bibinfo  {journal} {Proceedings of the National Academy of Sciences}\ }\textbf {\bibinfo {volume} {120}},\ \bibinfo {pages} {e2221048120} (\bibinfo {year} {2023}{\natexlab{a}})}\BibitemShut {NoStop}%
\bibitem [{\citenamefont {Cao}\ \emph {et~al.}(2023)\citenamefont {Cao}, \citenamefont {Qiu}, \citenamefont {Kalin},\ and\ \citenamefont {Huang}}]{Cao2023}%
  \BibitemOpen
  \bibfield  {author} {\bibinfo {author} {\bibfnamefont {S.}~\bibnamefont {Cao}}, \bibinfo {author} {\bibfnamefont {Y.}~\bibnamefont {Qiu}}, \bibinfo {author} {\bibfnamefont {M.~L.}\ \bibnamefont {Kalin}}, \ and\ \bibinfo {author} {\bibfnamefont {X.}~\bibnamefont {Huang}},\ }\bibfield  {title} {\enquote {\bibinfo {title} {Integrative generalized master equation: A method to study long-timescale biomolecular dynamics via the integrals of memory kernels},}\ }\href {\doibase 10.1063/5.0167287} {\bibfield  {journal} {\bibinfo  {journal} {The Journal of Chemical Physics}\ }\textbf {\bibinfo {volume} {159}},\ \bibinfo {pages} {134106} (\bibinfo {year} {2023})}\BibitemShut {NoStop}%
\bibitem [{\citenamefont {Lange}\ and\ \citenamefont {Grubmüller}(2006)}]{Lange2006}%
  \BibitemOpen
  \bibfield  {author} {\bibinfo {author} {\bibfnamefont {O.~F.}\ \bibnamefont {Lange}}\ and\ \bibinfo {author} {\bibfnamefont {H.}~\bibnamefont {Grubmüller}},\ }\bibfield  {title} {\enquote {\bibinfo {title} {Collective langevin dynamics of conformational motions in proteins},}\ }\href {\doibase 10.1063/1.2199530} {\bibfield  {journal} {\bibinfo  {journal} {The Journal of Chemical Physics}\ }\textbf {\bibinfo {volume} {124}},\ \bibinfo {pages} {17} (\bibinfo {year} {2006})}\BibitemShut {NoStop}%
\bibitem [{\citenamefont {Carof}\ \emph {et~al.}(2014)\citenamefont {Carof}, \citenamefont {Marry}, \citenamefont {Salanne}, \citenamefont {Hansen}, \citenamefont {Turq},\ and\ \citenamefont {Rotenberg}}]{Carof2014a}%
  \BibitemOpen
  \bibfield  {author} {\bibinfo {author} {\bibfnamefont {A.}~\bibnamefont {Carof}}, \bibinfo {author} {\bibfnamefont {V.}~\bibnamefont {Marry}}, \bibinfo {author} {\bibfnamefont {M.}~\bibnamefont {Salanne}}, \bibinfo {author} {\bibfnamefont {J.~P.}\ \bibnamefont {Hansen}}, \bibinfo {author} {\bibfnamefont {P.}~\bibnamefont {Turq}}, \ and\ \bibinfo {author} {\bibfnamefont {B.}~\bibnamefont {Rotenberg}},\ }\bibfield  {title} {\enquote {\bibinfo {title} {Coarse graining the dynamics of nano-confined solutes: the case of ions in clays},}\ }\href {\doibase 10.1080/08927022.2013.840894} {\bibfield  {journal} {\bibinfo  {journal} {Molecular Simulation}\ }\textbf {\bibinfo {volume} {40}},\ \bibinfo {pages} {237--244} (\bibinfo {year} {2014})}\BibitemShut {NoStop}%
\bibitem [{\citenamefont {Lesnicki}\ \emph {et~al.}(2016)\citenamefont {Lesnicki}, \citenamefont {Vuilleumier}, \citenamefont {Carof},\ and\ \citenamefont {Rotenberg}}]{Lesnicki2016}%
  \BibitemOpen
  \bibfield  {author} {\bibinfo {author} {\bibfnamefont {D.}~\bibnamefont {Lesnicki}}, \bibinfo {author} {\bibfnamefont {R.}~\bibnamefont {Vuilleumier}}, \bibinfo {author} {\bibfnamefont {A.}~\bibnamefont {Carof}}, \ and\ \bibinfo {author} {\bibfnamefont {B.}~\bibnamefont {Rotenberg}},\ }\bibfield  {title} {\enquote {\bibinfo {title} {Molecular hydrodynamics from memory kernels},}\ }\href {\doibase 10.1103/PHYSREVLETT.116.147804} {\bibfield  {journal} {\bibinfo  {journal} {Physical Review Letters}\ }\textbf {\bibinfo {volume} {116}},\ \bibinfo {pages} {147804} (\bibinfo {year} {2016})}\BibitemShut {NoStop}%
\bibitem [{\citenamefont {Kowalik}\ \emph {et~al.}(2019)\citenamefont {Kowalik}, \citenamefont {Daldrop}, \citenamefont {Kappler}, \citenamefont {Schulz}, \citenamefont {Schlaich},\ and\ \citenamefont {Netz}}]{Kowalik2019}%
  \BibitemOpen
  \bibfield  {author} {\bibinfo {author} {\bibfnamefont {B.}~\bibnamefont {Kowalik}}, \bibinfo {author} {\bibfnamefont {J.~O.}\ \bibnamefont {Daldrop}}, \bibinfo {author} {\bibfnamefont {J.}~\bibnamefont {Kappler}}, \bibinfo {author} {\bibfnamefont {J.~C.}\ \bibnamefont {Schulz}}, \bibinfo {author} {\bibfnamefont {A.}~\bibnamefont {Schlaich}}, \ and\ \bibinfo {author} {\bibfnamefont {R.~R.}\ \bibnamefont {Netz}},\ }\bibfield  {title} {\enquote {\bibinfo {title} {Memory-kernel extraction for different molecular solutes in solvents of varying viscosity in confinement},}\ }\href {\doibase 10.1103/PHYSREVE.100.012126} {\bibfield  {journal} {\bibinfo  {journal} {Physical Review E}\ }\textbf {\bibinfo {volume} {100}},\ \bibinfo {pages} {012126} (\bibinfo {year} {2019})}\BibitemShut {NoStop}%
\bibitem [{\citenamefont {Brünig}\ \emph {et~al.}(2022)\citenamefont {Brünig}, \citenamefont {Geburtig}, \citenamefont {Canal}, \citenamefont {Kappler},\ and\ \citenamefont {Netz}}]{Bruenig2022}%
  \BibitemOpen
  \bibfield  {author} {\bibinfo {author} {\bibfnamefont {F.~N.}\ \bibnamefont {Brünig}}, \bibinfo {author} {\bibfnamefont {O.}~\bibnamefont {Geburtig}}, \bibinfo {author} {\bibfnamefont {A.~V.}\ \bibnamefont {Canal}}, \bibinfo {author} {\bibfnamefont {J.}~\bibnamefont {Kappler}}, \ and\ \bibinfo {author} {\bibfnamefont {R.~R.}\ \bibnamefont {Netz}},\ }\bibfield  {title} {\enquote {\bibinfo {title} {Time-dependent friction effects on vibrational infrared frequencies and line shapes of liquid water},}\ }\href {\doibase 10.1021/ACS.JPCB.1C09481} {\bibfield  {journal} {\bibinfo  {journal} {The Journal of Physical Chemistry B}\ }\textbf {\bibinfo {volume} {126}},\ \bibinfo {pages} {1579--1589} (\bibinfo {year} {2022})}\BibitemShut {NoStop}%
\bibitem [{\citenamefont {Vroylandt}\ \emph {et~al.}(2022)\citenamefont {Vroylandt}, \citenamefont {Goudenège}, \citenamefont {Monmarché}, \citenamefont {Pietrucci},\ and\ \citenamefont {Rotenberg}}]{Vroylandt2022a}%
  \BibitemOpen
  \bibfield  {author} {\bibinfo {author} {\bibfnamefont {H.}~\bibnamefont {Vroylandt}}, \bibinfo {author} {\bibfnamefont {L.}~\bibnamefont {Goudenège}}, \bibinfo {author} {\bibfnamefont {P.}~\bibnamefont {Monmarché}}, \bibinfo {author} {\bibfnamefont {F.}~\bibnamefont {Pietrucci}}, \ and\ \bibinfo {author} {\bibfnamefont {B.}~\bibnamefont {Rotenberg}},\ }\bibfield  {title} {\enquote {\bibinfo {title} {Likelihood-based non-markovian models from molecular dynamics},}\ }\href {\doibase 10.1073/pnas.2117586119} {\bibfield  {journal} {\bibinfo  {journal} {Proceedings of the National Academy of Sciences}\ }\textbf {\bibinfo {volume} {119}},\ \bibinfo {pages} {e2117586119} (\bibinfo {year} {2022})}\BibitemShut {NoStop}%
\bibitem [{\citenamefont {Vroylandt}\ and\ \citenamefont {Monmarché}(2022)}]{Vroylandt2022b}%
  \BibitemOpen
  \bibfield  {author} {\bibinfo {author} {\bibfnamefont {H.}~\bibnamefont {Vroylandt}}\ and\ \bibinfo {author} {\bibfnamefont {P.}~\bibnamefont {Monmarché}},\ }\bibfield  {title} {\enquote {\bibinfo {title} {Position-dependent memory kernel in generalized langevin equations: Theory and numerical estimation},}\ }\href {\doibase 10.1063/5.0094566} {\bibfield  {journal} {\bibinfo  {journal} {Journal of Chemical Physics}\ }\textbf {\bibinfo {volume} {156}},\ \bibinfo {pages} {244105} (\bibinfo {year} {2022})}\BibitemShut {NoStop}%
\bibitem [{\citenamefont {Kiefer}\ \emph {et~al.}(2025{\natexlab{a}})\citenamefont {Kiefer}, \citenamefont {Héry}, \citenamefont {Tepper}, \citenamefont {Dalton}, \citenamefont {Ayaz},\ and\ \citenamefont {Netz}}]{KieferArchive}%
  \BibitemOpen
  \bibfield  {author} {\bibinfo {author} {\bibfnamefont {H.}~\bibnamefont {Kiefer}}, \bibinfo {author} {\bibfnamefont {B.~J.~A.}\ \bibnamefont {Héry}}, \bibinfo {author} {\bibfnamefont {L.}~\bibnamefont {Tepper}}, \bibinfo {author} {\bibfnamefont {B.~A.}\ \bibnamefont {Dalton}}, \bibinfo {author} {\bibfnamefont {C.}~\bibnamefont {Ayaz}}, \ and\ \bibinfo {author} {\bibfnamefont {R.~R.}\ \bibnamefont {Netz}},\ }\bibfield  {title} {\enquote {\bibinfo {title} {Analysis and simulation of generalized langevin equations with non-gaussian orthogonal forces},}\ }\href {https://arxiv.org/pdf/2505.15665} {\bibfield  {journal} {\bibinfo  {journal} {arXiv:2505.15665 [physics.comp-ph]}\ } (\bibinfo {year} {2025}{\natexlab{a}})}\BibitemShut {NoStop}%
\bibitem [{\citenamefont {Kiefer}\ \emph {et~al.}(2025{\natexlab{b}})\citenamefont {Kiefer}, \citenamefont {Furtel}, \citenamefont {Ayaz}, \citenamefont {Klimek}, \citenamefont {Daldrop},\ and\ \citenamefont {Netz}}]{Kiefer2025}%
  \BibitemOpen
  \bibfield  {author} {\bibinfo {author} {\bibfnamefont {H.}~\bibnamefont {Kiefer}}, \bibinfo {author} {\bibfnamefont {D.}~\bibnamefont {Furtel}}, \bibinfo {author} {\bibfnamefont {C.}~\bibnamefont {Ayaz}}, \bibinfo {author} {\bibfnamefont {A.}~\bibnamefont {Klimek}}, \bibinfo {author} {\bibfnamefont {J.~O.}\ \bibnamefont {Daldrop}}, \ and\ \bibinfo {author} {\bibfnamefont {R.~R.}\ \bibnamefont {Netz}},\ }\bibfield  {title} {\enquote {\bibinfo {title} {Prediction of weather and financial time-series data via a hamiltonian-based filter-projection approach},}\ }\href {\doibase 10.1016/j.newton.2025.100138} {\bibfield  {journal} {\bibinfo  {journal} {Newton}\ }\textbf {\bibinfo {volume} {1}},\ \bibinfo {pages} {100138} (\bibinfo {year} {2025}{\natexlab{b}})}\BibitemShut {NoStop}%
\bibitem [{Note3()}]{Note3}%
  \BibitemOpen
  \bibinfo {note} {There is a case where the memory kernel fits a model form that can be extrapolated,\cite {Klimek2025} but in general long-timescales in the memory kernel cannot easily be inferred from short-time data, especially in the presence of noise.}\BibitemShut {Stop}%
\bibitem [{Note4()}]{Note4}%
  \BibitemOpen
  \bibinfo {note} {This can be seen as the opposite of a method like DMD that takes the full configurational space and extracts the high/infinite dimensional (Markovian) propagator by performing a dimensionality reduction, effectively approaching the minimal space from above by rank reduction.\cite {Schmid2022}}\BibitemShut {NoStop}%
\bibitem [{\citenamefont {Gaigeot}\ and\ \citenamefont {Sprik}(2003)}]{Gaigeot2003}%
  \BibitemOpen
  \bibfield  {author} {\bibinfo {author} {\bibfnamefont {M.~P.}\ \bibnamefont {Gaigeot}}\ and\ \bibinfo {author} {\bibfnamefont {M.}~\bibnamefont {Sprik}},\ }\bibfield  {title} {\enquote {\bibinfo {title} {Ab initio molecular dynamics computation of the infrared spectrum of aqueous uracil},}\ }\href {\doibase 10.1021/JP034788U} {\bibfield  {journal} {\bibinfo  {journal} {Journal of Physical Chemistry B}\ }\textbf {\bibinfo {volume} {107}},\ \bibinfo {pages} {10344--10358} (\bibinfo {year} {2003})}\BibitemShut {NoStop}%
\bibitem [{\citenamefont {Auer}\ and\ \citenamefont {Skinner}(2008)}]{Auer2008}%
  \BibitemOpen
  \bibfield  {author} {\bibinfo {author} {\bibfnamefont {B.~M.}\ \bibnamefont {Auer}}\ and\ \bibinfo {author} {\bibfnamefont {J.~L.}\ \bibnamefont {Skinner}},\ }\bibfield  {title} {\enquote {\bibinfo {title} {Vibrational sum-frequency spectroscopy of the liquid/vapor interface for dilute hod in d2o},}\ }\href {\doibase 10.1063/1.3012568} {\bibfield  {journal} {\bibinfo  {journal} {The Journal of Chemical Physics}\ }\textbf {\bibinfo {volume} {129}},\ \bibinfo {pages} {214705} (\bibinfo {year} {2008})}\BibitemShut {NoStop}%
\bibitem [{\citenamefont {Stiopkin}\ \emph {et~al.}(2011)\citenamefont {Stiopkin}, \citenamefont {Weeraman}, \citenamefont {Pieniazek}, \citenamefont {Shalhout}, \citenamefont {Skinner},\ and\ \citenamefont {Benderskii}}]{Stiopkin2011}%
  \BibitemOpen
  \bibfield  {author} {\bibinfo {author} {\bibfnamefont {I.~V.}\ \bibnamefont {Stiopkin}}, \bibinfo {author} {\bibfnamefont {C.}~\bibnamefont {Weeraman}}, \bibinfo {author} {\bibfnamefont {P.~A.}\ \bibnamefont {Pieniazek}}, \bibinfo {author} {\bibfnamefont {F.~Y.}\ \bibnamefont {Shalhout}}, \bibinfo {author} {\bibfnamefont {J.~L.}\ \bibnamefont {Skinner}}, \ and\ \bibinfo {author} {\bibfnamefont {A.~V.}\ \bibnamefont {Benderskii}},\ }\bibfield  {title} {\enquote {\bibinfo {title} {Hydrogen bonding at the water surface revealed by isotopic dilution spectroscopy},}\ }\href {\doibase 10.1038/NATURE10173} {\bibfield  {journal} {\bibinfo  {journal} {Nature}\ }\textbf {\bibinfo {volume} {474}},\ \bibinfo {pages} {192--195} (\bibinfo {year} {2011})}\BibitemShut {NoStop}%
\bibitem [{\citenamefont {Sulpizi}\ \emph {et~al.}(2012)\citenamefont {Sulpizi}, \citenamefont {Salanne}, \citenamefont {Sprik},\ and\ \citenamefont {Gaigeot}}]{Sulpizi2012}%
  \BibitemOpen
  \bibfield  {author} {\bibinfo {author} {\bibfnamefont {M.}~\bibnamefont {Sulpizi}}, \bibinfo {author} {\bibfnamefont {M.}~\bibnamefont {Salanne}}, \bibinfo {author} {\bibfnamefont {M.}~\bibnamefont {Sprik}}, \ and\ \bibinfo {author} {\bibfnamefont {M.~P.}\ \bibnamefont {Gaigeot}},\ }\bibfield  {title} {\enquote {\bibinfo {title} {Vibrational sum frequency generation spectroscopy of the water liquid–vapor interface from density functional theory-based molecular dynamics simulations},}\ }\href {\doibase 10.1021/JZ301858G} {\bibfield  {journal} {\bibinfo  {journal} {Journal of Physical Chemistry Letters}\ }\textbf {\bibinfo {volume} {4}},\ \bibinfo {pages} {83--87} (\bibinfo {year} {2012})}\BibitemShut {NoStop}%
\bibitem [{\citenamefont {Nagata}\ \emph {et~al.}(2013)\citenamefont {Nagata}, \citenamefont {Hsieh}, \citenamefont {Hasegawa}, \citenamefont {Voll}, \citenamefont {Backus},\ and\ \citenamefont {Bonn}}]{Nagata2013}%
  \BibitemOpen
  \bibfield  {author} {\bibinfo {author} {\bibfnamefont {Y.}~\bibnamefont {Nagata}}, \bibinfo {author} {\bibfnamefont {C.~S.}\ \bibnamefont {Hsieh}}, \bibinfo {author} {\bibfnamefont {T.}~\bibnamefont {Hasegawa}}, \bibinfo {author} {\bibfnamefont {J.}~\bibnamefont {Voll}}, \bibinfo {author} {\bibfnamefont {E.~H.}\ \bibnamefont {Backus}}, \ and\ \bibinfo {author} {\bibfnamefont {M.}~\bibnamefont {Bonn}},\ }\bibfield  {title} {\enquote {\bibinfo {title} {Water bending mode at the water–vapor interface probed by sum-frequency generation spectroscopy: A combined molecular dynamics simulation and experimental study},}\ }\href {\doibase 10.1021/JZ400683V} {\bibfield  {journal} {\bibinfo  {journal} {Journal of Physical Chemistry Letters}\ }\textbf {\bibinfo {volume} {4}},\ \bibinfo {pages} {1872--1877} (\bibinfo {year} {2013})}\BibitemShut {NoStop}%
\bibitem [{\citenamefont {Ohto}\ \emph {et~al.}(2015)\citenamefont {Ohto}, \citenamefont {Usui}, \citenamefont {Hasegawa}, \citenamefont {Bonn},\ and\ \citenamefont {Nagata}}]{Ohto2015}%
  \BibitemOpen
  \bibfield  {author} {\bibinfo {author} {\bibfnamefont {T.}~\bibnamefont {Ohto}}, \bibinfo {author} {\bibfnamefont {K.}~\bibnamefont {Usui}}, \bibinfo {author} {\bibfnamefont {T.}~\bibnamefont {Hasegawa}}, \bibinfo {author} {\bibfnamefont {M.}~\bibnamefont {Bonn}}, \ and\ \bibinfo {author} {\bibfnamefont {Y.}~\bibnamefont {Nagata}},\ }\bibfield  {title} {\enquote {\bibinfo {title} {Toward ab initio molecular dynamics modeling for sum-frequency generation spectra; an efficient algorithm based on surface-specific velocity-velocity correlation function},}\ }\href {\doibase 10.1063/1.4931106} {\bibfield  {journal} {\bibinfo  {journal} {The Journal of Chemical Physics}\ }\textbf {\bibinfo {volume} {143}} (\bibinfo {year} {2015}),\ 10.1063/1.4931106}\BibitemShut {NoStop}%
\bibitem [{\citenamefont {Khatib}\ \emph {et~al.}(2016)\citenamefont {Khatib}, \citenamefont {Backus}, \citenamefont {Bonn}, \citenamefont {Perez-Haro}, \citenamefont {Gaigeot},\ and\ \citenamefont {Sulpizi}}]{Khatib2016}%
  \BibitemOpen
  \bibfield  {author} {\bibinfo {author} {\bibfnamefont {R.}~\bibnamefont {Khatib}}, \bibinfo {author} {\bibfnamefont {E.~H.}\ \bibnamefont {Backus}}, \bibinfo {author} {\bibfnamefont {M.}~\bibnamefont {Bonn}}, \bibinfo {author} {\bibfnamefont {M.~J.}\ \bibnamefont {Perez-Haro}}, \bibinfo {author} {\bibfnamefont {M.~P.}\ \bibnamefont {Gaigeot}}, \ and\ \bibinfo {author} {\bibfnamefont {M.}~\bibnamefont {Sulpizi}},\ }\bibfield  {title} {\enquote {\bibinfo {title} {Water orientation and hydrogen-bond structure at the fluorite/water interface},}\ }\href {\doibase 10.1038/SREP24287} {\bibfield  {journal} {\bibinfo  {journal} {Scientific Reports}\ }\textbf {\bibinfo {volume} {6}},\ \bibinfo {pages} {1--10} (\bibinfo {year} {2016})}\BibitemShut {NoStop}%
\bibitem [{\citenamefont {Khatib}\ and\ \citenamefont {Sulpizi}(2017)}]{Khatib2017}%
  \BibitemOpen
  \bibfield  {author} {\bibinfo {author} {\bibfnamefont {R.}~\bibnamefont {Khatib}}\ and\ \bibinfo {author} {\bibfnamefont {M.}~\bibnamefont {Sulpizi}},\ }\bibfield  {title} {\enquote {\bibinfo {title} {Sum frequency generation spectra from velocity–velocity correlation functions},}\ }\href {\doibase 10.1021/ACS.JPCLETT.7B00207} {\bibfield  {journal} {\bibinfo  {journal} {Journal of Physical Chemistry Letters}\ }\textbf {\bibinfo {volume} {8}},\ \bibinfo {pages} {1310--1314} (\bibinfo {year} {2017})}\BibitemShut {NoStop}%
\bibitem [{\citenamefont {Moberg}, \citenamefont {Straight},\ and\ \citenamefont {Paesani}(2018)}]{Moberg2018}%
  \BibitemOpen
  \bibfield  {author} {\bibinfo {author} {\bibfnamefont {D.~R.}\ \bibnamefont {Moberg}}, \bibinfo {author} {\bibfnamefont {S.~C.}\ \bibnamefont {Straight}}, \ and\ \bibinfo {author} {\bibfnamefont {F.}~\bibnamefont {Paesani}},\ }\bibfield  {title} {\enquote {\bibinfo {title} {Temperature dependence of the air/water interface revealed by polarization sensitive sum-frequency generation spectroscopy},}\ }\href {\doibase 10.1021/ACS.JPCB.8B01726} {\bibfield  {journal} {\bibinfo  {journal} {The Journal of Physical Chemistry B}\ }\textbf {\bibinfo {volume} {122}},\ \bibinfo {pages} {4356--4365} (\bibinfo {year} {2018})}\BibitemShut {NoStop}%
\bibitem [{\citenamefont {Litman}\ \emph {et~al.}(2023)\citenamefont {Litman}, \citenamefont {Lan}, \citenamefont {Nagata},\ and\ \citenamefont {Wilkins}}]{Litman2023}%
  \BibitemOpen
  \bibfield  {author} {\bibinfo {author} {\bibfnamefont {Y.}~\bibnamefont {Litman}}, \bibinfo {author} {\bibfnamefont {J.}~\bibnamefont {Lan}}, \bibinfo {author} {\bibfnamefont {Y.}~\bibnamefont {Nagata}}, \ and\ \bibinfo {author} {\bibfnamefont {D.~M.}\ \bibnamefont {Wilkins}},\ }\bibfield  {title} {\enquote {\bibinfo {title} {Fully first-principles surface spectroscopy with machine learning},}\ }\href {\doibase 10.1021/ACS.JPCLETT.3C01989} {\bibfield  {journal} {\bibinfo  {journal} {The Journal of Physical Chemistry Letters}\ }\textbf {\bibinfo {volume} {14}},\ \bibinfo {pages} {8175--8182} (\bibinfo {year} {2023})}\BibitemShut {NoStop}%
\bibitem [{\citenamefont {Jana}\ \emph {et~al.}(2024)\citenamefont {Jana}, \citenamefont {Shepherd}, \citenamefont {Litman},\ and\ \citenamefont {Wilkins}}]{Jana2024}%
  \BibitemOpen
  \bibfield  {author} {\bibinfo {author} {\bibfnamefont {A.}~\bibnamefont {Jana}}, \bibinfo {author} {\bibfnamefont {S.}~\bibnamefont {Shepherd}}, \bibinfo {author} {\bibfnamefont {Y.}~\bibnamefont {Litman}}, \ and\ \bibinfo {author} {\bibfnamefont {D.~M.}\ \bibnamefont {Wilkins}},\ }\bibfield  {title} {\enquote {\bibinfo {title} {Learning electronic polarizations in aqueous systems},}\ }\href {\doibase 10.1021/ACS.JCIM.4C00421} {\bibfield  {journal} {\bibinfo  {journal} {Journal of Chemical Information and Modeling}\ }\textbf {\bibinfo {volume} {64}},\ \bibinfo {pages} {4426--4435} (\bibinfo {year} {2024})}\BibitemShut {NoStop}%
\bibitem [{\citenamefont {Fellows}\ \emph {et~al.}(2024{\natexlab{a}})\citenamefont {Fellows}, \citenamefont {Álvaro Díaz~Duque}, \citenamefont {Balos}, \citenamefont {Lehmann}, \citenamefont {Netz}, \citenamefont {Wolf},\ and\ \citenamefont {Thämer}}]{Fellows2024}%
  \BibitemOpen
  \bibfield  {author} {\bibinfo {author} {\bibfnamefont {A.~P.}\ \bibnamefont {Fellows}}, \bibinfo {author} {\bibnamefont {Álvaro Díaz~Duque}}, \bibinfo {author} {\bibfnamefont {V.}~\bibnamefont {Balos}}, \bibinfo {author} {\bibfnamefont {L.}~\bibnamefont {Lehmann}}, \bibinfo {author} {\bibfnamefont {R.~R.}\ \bibnamefont {Netz}}, \bibinfo {author} {\bibfnamefont {M.}~\bibnamefont {Wolf}}, \ and\ \bibinfo {author} {\bibfnamefont {M.}~\bibnamefont {Thämer}},\ }\bibfield  {title} {\enquote {\bibinfo {title} {Sum-frequency generation spectroscopy of aqueous interfaces: The role of depth and its impact on spectral interpretation},}\ }\href {\doibase 10.1021/ACS.JPCC.4C06650} {\bibfield  {journal} {\bibinfo  {journal} {Journal of Physical Chemistry C}\ }\textbf {\bibinfo {volume} {128}},\ \bibinfo {pages} {20733--20750} (\bibinfo {year} {2024}{\natexlab{a}})}\BibitemShut {NoStop}%
\bibitem [{\citenamefont {Fellows}\ \emph {et~al.}(2024{\natexlab{b}})\citenamefont {Fellows}, \citenamefont {Álvaro Díaz~Duque}, \citenamefont {Balos}, \citenamefont {Lehmann}, \citenamefont {Netz}, \citenamefont {Wolf},\ and\ \citenamefont {Thämer}}]{Fellows2024b}%
  \BibitemOpen
  \bibfield  {author} {\bibinfo {author} {\bibfnamefont {A.~P.}\ \bibnamefont {Fellows}}, \bibinfo {author} {\bibnamefont {Álvaro Díaz~Duque}}, \bibinfo {author} {\bibfnamefont {V.}~\bibnamefont {Balos}}, \bibinfo {author} {\bibfnamefont {L.}~\bibnamefont {Lehmann}}, \bibinfo {author} {\bibfnamefont {R.~R.}\ \bibnamefont {Netz}}, \bibinfo {author} {\bibfnamefont {M.}~\bibnamefont {Wolf}}, \ and\ \bibinfo {author} {\bibfnamefont {M.}~\bibnamefont {Thämer}},\ }\bibfield  {title} {\enquote {\bibinfo {title} {How thick is the air-water interface?---a direct experimental measurement of the decay length of the interfacial structural anisotropy},}\ }\href {\doibase 10.1021/ACS.LANGMUIR.4C02571} {\bibfield  {journal} {\bibinfo  {journal} {Langmuir}\ }\textbf {\bibinfo {volume} {40}},\ \bibinfo {pages} {18760--18772} (\bibinfo {year} {2024}{\natexlab{b}})}\BibitemShut {NoStop}%
\bibitem [{\citenamefont {Morita}(2018)}]{BookMorita}%
  \BibitemOpen
  \bibfield  {author} {\bibinfo {author} {\bibfnamefont {A.}~\bibnamefont {Morita}},\ }\href {\doibase 10.1007/978-981-13-1607-4} {\emph {\bibinfo {title} {Theory of Sum Frequency Generation Spectroscopy}}},\ \bibinfo {edition} {1st}\ ed.,\ Vol.~\bibinfo {volume} {97}\ (\bibinfo  {publisher} {Springer Nature Singapore},\ \bibinfo {year} {2018})\BibitemShut {NoStop}%
\bibitem [{\citenamefont {Shen}(2002)}]{BookShen}%
  \BibitemOpen
  \bibfield  {author} {\bibinfo {author} {\bibfnamefont {Y.~R.}\ \bibnamefont {Shen}},\ }\href {http://www.amazon.com/exec/obidos/redirect?tag=citeulike07-20\&path=ASIN/0471430803} {\emph {\bibinfo {title} {The Principles of Nonlinear Optics}}},\ \bibinfo {edition} {1st}\ ed.\ (\bibinfo  {publisher} {Wiley-Interscience},\ \bibinfo {year} {2002})\BibitemShut {NoStop}%
\bibitem [{\citenamefont {Morita}(2006)}]{Morita2006}%
  \BibitemOpen
  \bibfield  {author} {\bibinfo {author} {\bibfnamefont {A.}~\bibnamefont {Morita}},\ }\bibfield  {title} {\enquote {\bibinfo {title} {Improved computation of sum frequency generation spectrum of the surface of water},}\ }\href {\doibase 10.1021/JP058155M} {\bibfield  {journal} {\bibinfo  {journal} {Journal of Physical Chemistry B}\ }\textbf {\bibinfo {volume} {110}},\ \bibinfo {pages} {3158--3163} (\bibinfo {year} {2006})}\BibitemShut {NoStop}%
\bibitem [{\citenamefont {Berendsen}, \citenamefont {Grigera},\ and\ \citenamefont {Straatsma}(1987)}]{SPCE}%
  \BibitemOpen
  \bibfield  {author} {\bibinfo {author} {\bibfnamefont {H.~J.}\ \bibnamefont {Berendsen}}, \bibinfo {author} {\bibfnamefont {J.~R.}\ \bibnamefont {Grigera}}, \ and\ \bibinfo {author} {\bibfnamefont {T.~P.}\ \bibnamefont {Straatsma}},\ }\bibfield  {title} {\enquote {\bibinfo {title} {The missing term in effective pair potentials},}\ }\href {\doibase 10.1021/J100308A038} {\bibfield  {journal} {\bibinfo  {journal} {Journal of Physical Chemistry}\ }\textbf {\bibinfo {volume} {91}},\ \bibinfo {pages} {6269--6271} (\bibinfo {year} {1987})}\BibitemShut {NoStop}%
\bibitem [{\citenamefont {Abraham}\ \emph {et~al.}(2015)\citenamefont {Abraham}, \citenamefont {Murtola}, \citenamefont {Schulz}, \citenamefont {Páll}, \citenamefont {Smith}, \citenamefont {Hess},\ and\ \citenamefont {Lindah}}]{GROMACS}%
  \BibitemOpen
  \bibfield  {author} {\bibinfo {author} {\bibfnamefont {M.~J.}\ \bibnamefont {Abraham}}, \bibinfo {author} {\bibfnamefont {T.}~\bibnamefont {Murtola}}, \bibinfo {author} {\bibfnamefont {R.}~\bibnamefont {Schulz}}, \bibinfo {author} {\bibfnamefont {S.}~\bibnamefont {Páll}}, \bibinfo {author} {\bibfnamefont {J.~C.}\ \bibnamefont {Smith}}, \bibinfo {author} {\bibfnamefont {B.}~\bibnamefont {Hess}}, \ and\ \bibinfo {author} {\bibfnamefont {E.}~\bibnamefont {Lindah}},\ }\bibfield  {title} {\enquote {\bibinfo {title} {Gromacs: High performance molecular simulations through multi-level parallelism from laptops to supercomputers},}\ }\href {\doibase 10.1016/J.SOFTX.2015.06.001} {\bibfield  {journal} {\bibinfo  {journal} {SoftwareX}\ }\textbf {\bibinfo {volume} {1-2}},\ \bibinfo {pages} {19--25} (\bibinfo {year} {2015})}\BibitemShut {NoStop}%
\bibitem [{\citenamefont {VandeVondele}\ \emph {et~al.}(2005)\citenamefont {VandeVondele}, \citenamefont {Krack}, \citenamefont {Mohamed}, \citenamefont {Parrinello}, \citenamefont {Chassaing},\ and\ \citenamefont {Hutter}}]{VandeVondele:2005ge}%
  \BibitemOpen
  \bibfield  {author} {\bibinfo {author} {\bibfnamefont {J.}~\bibnamefont {VandeVondele}}, \bibinfo {author} {\bibfnamefont {M.}~\bibnamefont {Krack}}, \bibinfo {author} {\bibfnamefont {F.}~\bibnamefont {Mohamed}}, \bibinfo {author} {\bibfnamefont {M.}~\bibnamefont {Parrinello}}, \bibinfo {author} {\bibfnamefont {T.}~\bibnamefont {Chassaing}}, \ and\ \bibinfo {author} {\bibfnamefont {J.}~\bibnamefont {Hutter}},\ }\bibfield  {title} {\enquote {\bibinfo {title} {Quickstep: Fast and accurate density functional calculations using a mixed gaussian and plane waves approach},}\ }\href {\doibase 10.1016/j.cpc.2004.12.014} {\bibfield  {journal} {\bibinfo  {journal} {Computer Physics Communications}\ }\textbf {\bibinfo {volume} {167}},\ \bibinfo {pages} {103--128} (\bibinfo {year} {2005})}\BibitemShut {NoStop}%
\bibitem [{\citenamefont {VandeVondele}\ and\ \citenamefont {Hutter}(2003)}]{VandeVondele:2003ue}%
  \BibitemOpen
  \bibfield  {author} {\bibinfo {author} {\bibfnamefont {J.}~\bibnamefont {VandeVondele}}\ and\ \bibinfo {author} {\bibfnamefont {J.}~\bibnamefont {Hutter}},\ }\bibfield  {title} {\enquote {\bibinfo {title} {An efficient orbital transformation method for electronic structure calculations},}\ }\href {\doibase 10.1063/1.1543154} {\bibfield  {journal} {\bibinfo  {journal} {The Journal of Chemical Physics}\ }\textbf {\bibinfo {volume} {118}},\ \bibinfo {pages} {4365} (\bibinfo {year} {2003})}\BibitemShut {NoStop}%
\bibitem [{\citenamefont {Perdew}, \citenamefont {Burke},\ and\ \citenamefont {Ernzerhof}(1996)}]{Perdew1996}%
  \BibitemOpen
  \bibfield  {author} {\bibinfo {author} {\bibfnamefont {J.~P.}\ \bibnamefont {Perdew}}, \bibinfo {author} {\bibfnamefont {K.}~\bibnamefont {Burke}}, \ and\ \bibinfo {author} {\bibfnamefont {M.}~\bibnamefont {Ernzerhof}},\ }\bibfield  {title} {\enquote {\bibinfo {title} {Generalized gradient approximation made simple},}\ }\href {\doibase 10.1103/PhysRevLett.77.3865} {\bibfield  {journal} {\bibinfo  {journal} {Physical Review Letters}\ }\textbf {\bibinfo {volume} {77}},\ \bibinfo {pages} {3865--3868} (\bibinfo {year} {1996})}\BibitemShut {NoStop}%
\bibitem [{\citenamefont {Grimme}\ \emph {et~al.}(2010)\citenamefont {Grimme}, \citenamefont {Antony}, \citenamefont {Ehrlich},\ and\ \citenamefont {Krieg}}]{GrimmeD3}%
  \BibitemOpen
  \bibfield  {author} {\bibinfo {author} {\bibfnamefont {S.}~\bibnamefont {Grimme}}, \bibinfo {author} {\bibfnamefont {J.}~\bibnamefont {Antony}}, \bibinfo {author} {\bibfnamefont {S.}~\bibnamefont {Ehrlich}}, \ and\ \bibinfo {author} {\bibfnamefont {H.}~\bibnamefont {Krieg}},\ }\bibfield  {title} {\enquote {\bibinfo {title} {A consistent and accurate ab initio parametrization of density functional dispersion correction (dft-d) for the 94 elements h-pu},}\ }\href {\doibase 10.1063/1.1564060} {\bibfield  {journal} {\bibinfo  {journal} {The Journal of Chemical Physics}\ }\textbf {\bibinfo {volume} {132}},\ \bibinfo {pages} {154104--1007} (\bibinfo {year} {2010})}\BibitemShut {NoStop}%
\bibitem [{\citenamefont {Hutter}\ \emph {et~al.}(2014)\citenamefont {Hutter}, \citenamefont {Iannuzzi}, \citenamefont {Schiffmann},\ and\ \citenamefont {VandeVondele}}]{CP2K}%
  \BibitemOpen
  \bibfield  {author} {\bibinfo {author} {\bibfnamefont {J.}~\bibnamefont {Hutter}}, \bibinfo {author} {\bibfnamefont {M.}~\bibnamefont {Iannuzzi}}, \bibinfo {author} {\bibfnamefont {F.}~\bibnamefont {Schiffmann}}, \ and\ \bibinfo {author} {\bibfnamefont {J.}~\bibnamefont {VandeVondele}},\ }\bibfield  {title} {\enquote {\bibinfo {title} {Cp2k: Atomistic simulations of condensed matter systems},}\ }\href {\doibase 10.1002/wcms.1159} {\bibfield  {journal} {\bibinfo  {journal} {Wiley Interdisciplinary Reviews: Computational Molecular Science}\ }\textbf {\bibinfo {volume} {4}},\ \bibinfo {pages} {15--25} (\bibinfo {year} {2014})}\BibitemShut {NoStop}%
\bibitem [{\citenamefont {Bussi}, \citenamefont {Donadio},\ and\ \citenamefont {Parrinello}(2007)}]{Bussi2007}%
  \BibitemOpen
  \bibfield  {author} {\bibinfo {author} {\bibfnamefont {G.}~\bibnamefont {Bussi}}, \bibinfo {author} {\bibfnamefont {D.}~\bibnamefont {Donadio}}, \ and\ \bibinfo {author} {\bibfnamefont {M.}~\bibnamefont {Parrinello}},\ }\bibfield  {title} {\enquote {\bibinfo {title} {Canonical sampling through velocity rescaling},}\ }\href {\doibase 10.1063/1.2408420} {\bibfield  {journal} {\bibinfo  {journal} {The Journal of Chemical Physics}\ }\textbf {\bibinfo {volume} {126}},\ \bibinfo {pages} {14101} (\bibinfo {year} {2007})}\BibitemShut {NoStop}%
\bibitem [{\citenamefont {Goedecker}, \citenamefont {Teter},\ and\ \citenamefont {Hutter}(1996)}]{Goedecker:1996ve}%
  \BibitemOpen
  \bibfield  {author} {\bibinfo {author} {\bibfnamefont {S.}~\bibnamefont {Goedecker}}, \bibinfo {author} {\bibfnamefont {M.}~\bibnamefont {Teter}}, \ and\ \bibinfo {author} {\bibfnamefont {J.}~\bibnamefont {Hutter}},\ }\bibfield  {title} {\enquote {\bibinfo {title} {Separable dual-space gaussian pseudopotentials},}\ }\href {\doibase 10.1103/PhysRevB.54.1703} {\bibfield  {journal} {\bibinfo  {journal} {Physical Review B}\ }\textbf {\bibinfo {volume} {54}},\ \bibinfo {pages} {1703--1710} (\bibinfo {year} {1996})}\BibitemShut {NoStop}%
\bibitem [{\citenamefont {Hartwigsen}, \citenamefont {Goedecker},\ and\ \citenamefont {Hutter}(1998)}]{Hartwigsen1998}%
  \BibitemOpen
  \bibfield  {author} {\bibinfo {author} {\bibfnamefont {C.}~\bibnamefont {Hartwigsen}}, \bibinfo {author} {\bibfnamefont {S.}~\bibnamefont {Goedecker}}, \ and\ \bibinfo {author} {\bibfnamefont {J.}~\bibnamefont {Hutter}},\ }\bibfield  {title} {\enquote {\bibinfo {title} {Relativistic separable dual-space gaussian pseudopotentials from h to rn},}\ }\href {\doibase 10.1103/PhysRevB.58.3641} {\bibfield  {journal} {\bibinfo  {journal} {Physical Review B}\ }\textbf {\bibinfo {volume} {58}},\ \bibinfo {pages} {3641} (\bibinfo {year} {1998})}\BibitemShut {NoStop}%
\bibitem [{\citenamefont {Brehm}\ \emph {et~al.}(2020)\citenamefont {Brehm}, \citenamefont {Thomas}, \citenamefont {Gehrke},\ and\ \citenamefont {Kirchner}}]{TRAVIS}%
  \BibitemOpen
  \bibfield  {author} {\bibinfo {author} {\bibfnamefont {M.}~\bibnamefont {Brehm}}, \bibinfo {author} {\bibfnamefont {M.}~\bibnamefont {Thomas}}, \bibinfo {author} {\bibfnamefont {S.}~\bibnamefont {Gehrke}}, \ and\ \bibinfo {author} {\bibfnamefont {B.}~\bibnamefont {Kirchner}},\ }\bibfield  {title} {\enquote {\bibinfo {title} {Travis—a free analyzer for trajectories from molecular simulation},}\ }\href {\doibase 10.1063/5.0005078} {\bibfield  {journal} {\bibinfo  {journal} {The Journal of Chemical Physics}\ }\textbf {\bibinfo {volume} {152}},\ \bibinfo {pages} {164105} (\bibinfo {year} {2020})}\BibitemShut {NoStop}%
\bibitem [{\citenamefont {Rycroft}(2009)}]{Rycroft2009}%
  \BibitemOpen
  \bibfield  {author} {\bibinfo {author} {\bibfnamefont {C.~H.}\ \bibnamefont {Rycroft}},\ }\bibfield  {title} {\enquote {\bibinfo {title} {Voro++: A three-dimensional voronoi cell library in c++},}\ }\href {\doibase 10.1063/1.3215722} {\bibfield  {journal} {\bibinfo  {journal} {Chaos: An Interdisciplinary Journal of Nonlinear Science}\ }\textbf {\bibinfo {volume} {19}},\ \bibinfo {pages} {041111} (\bibinfo {year} {2009})}\BibitemShut {NoStop}%
\bibitem [{\citenamefont {Thomas}, \citenamefont {Brehm},\ and\ \citenamefont {Kirchner}(2015)}]{Thomas2015}%
  \BibitemOpen
  \bibfield  {author} {\bibinfo {author} {\bibfnamefont {M.}~\bibnamefont {Thomas}}, \bibinfo {author} {\bibfnamefont {M.}~\bibnamefont {Brehm}}, \ and\ \bibinfo {author} {\bibfnamefont {B.}~\bibnamefont {Kirchner}},\ }\bibfield  {title} {\enquote {\bibinfo {title} {Voronoi dipole moments for the simulation of bulk phase vibrational spectra},}\ }\href {\doibase 10.1039/C4CP05272B} {\bibfield  {journal} {\bibinfo  {journal} {Physical Chemistry Chemical Physics}\ }\textbf {\bibinfo {volume} {17}},\ \bibinfo {pages} {3207--3213} (\bibinfo {year} {2015})}\BibitemShut {NoStop}%
\bibitem [{Note5()}]{Note5}%
  \BibitemOpen
  \bibinfo {note} {We would like to thank the first anonymous reviewer for pointing out the significance of neglecting the net molecular charges.}\BibitemShut {Stop}%
\bibitem [{\citenamefont {Bertie}\ and\ \citenamefont {Lan}(1996)}]{Bertie1996}%
  \BibitemOpen
  \bibfield  {author} {\bibinfo {author} {\bibfnamefont {J.~E.}\ \bibnamefont {Bertie}}\ and\ \bibinfo {author} {\bibfnamefont {Z.}~\bibnamefont {Lan}},\ }\bibfield  {title} {\enquote {\bibinfo {title} {Infrared intensities of liquids xx: The intensity of the oh stretching band of liquid water revisited, and the best current values of the optical constants of h2o(l) at 25°c between 15,000 and 1 cm-1},}\ }\href {\doibase 10.1366/0003702963905385} {\bibfield  {journal} {\bibinfo  {journal} {Applied Spectroscopy}\ }\textbf {\bibinfo {volume} {50}},\ \bibinfo {pages} {1047--1057} (\bibinfo {year} {1996})}\BibitemShut {NoStop}%
\bibitem [{\citenamefont {Nagata}\ and\ \citenamefont {Mukamel}(2010)}]{Nagata2010}%
  \BibitemOpen
  \bibfield  {author} {\bibinfo {author} {\bibfnamefont {Y.}~\bibnamefont {Nagata}}\ and\ \bibinfo {author} {\bibfnamefont {S.}~\bibnamefont {Mukamel}},\ }\bibfield  {title} {\enquote {\bibinfo {title} {Vibrational sum-frequency generation spectroscopy at the water/lipid interface: Molecular dynamics simulation study},}\ }\href {\doibase 10.1021/JA100508N} {\bibfield  {journal} {\bibinfo  {journal} {Journal of the American Chemical Society}\ }\textbf {\bibinfo {volume} {132}},\ \bibinfo {pages} {6434--6442} (\bibinfo {year} {2010})}\BibitemShut {NoStop}%
\bibitem [{\citenamefont {Cerrillo}\ and\ \citenamefont {Cao}(2014)}]{Cerrillo2014a}%
  \BibitemOpen
  \bibfield  {author} {\bibinfo {author} {\bibfnamefont {J.}~\bibnamefont {Cerrillo}}\ and\ \bibinfo {author} {\bibfnamefont {J.}~\bibnamefont {Cao}},\ }\bibfield  {title} {\enquote {\bibinfo {title} {Non-markovian dynamical maps: Numerical processing of open quantum trajectories},}\ }\href {\doibase 10.1103/PHYSREVLETT.112.110401} {\bibfield  {journal} {\bibinfo  {journal} {Physical Review Letters}\ }\textbf {\bibinfo {volume} {112}},\ \bibinfo {pages} {110401} (\bibinfo {year} {2014})}\BibitemShut {NoStop}%
\bibitem [{\citenamefont {Makri}\ \emph {et~al.}(2025)\citenamefont {Makri}, \citenamefont {Kundu}, \citenamefont {Cai},\ and\ \citenamefont {Wang}}]{Makri2025}%
  \BibitemOpen
  \bibfield  {author} {\bibinfo {author} {\bibfnamefont {N.}~\bibnamefont {Makri}}, \bibinfo {author} {\bibfnamefont {S.}~\bibnamefont {Kundu}}, \bibinfo {author} {\bibfnamefont {Z.}~\bibnamefont {Cai}}, \ and\ \bibinfo {author} {\bibfnamefont {G.}~\bibnamefont {Wang}},\ }\bibfield  {title} {\enquote {\bibinfo {title} {Comment on "unified framework for open quantum dynamics with memory"},}\ }\href {http://arxiv.org/abs/2410.08239} {\bibfield  {journal} {\bibinfo  {journal} {arXiv:2410.08239 [quant-ph]}\ }\textbf {\bibinfo {volume} {16}},\ \bibinfo {pages} {1--3} (\bibinfo {year} {2025})}\BibitemShut {NoStop}%
\bibitem [{\citenamefont {Peng}\ \emph {et~al.}(2025)\citenamefont {Peng}, \citenamefont {Ivander}, \citenamefont {Lindoy},\ and\ \citenamefont {Lee}}]{Peng2025}%
  \BibitemOpen
  \bibfield  {author} {\bibinfo {author} {\bibfnamefont {R.}~\bibnamefont {Peng}}, \bibinfo {author} {\bibfnamefont {F.}~\bibnamefont {Ivander}}, \bibinfo {author} {\bibfnamefont {L.~P.}\ \bibnamefont {Lindoy}}, \ and\ \bibinfo {author} {\bibfnamefont {J.}~\bibnamefont {Lee}},\ }\bibfield  {title} {\enquote {\bibinfo {title} {Addendum: Unified framework for open quantum dynamics with memory},}\ }\href {\doibase 10.1038/s41467-025-61825-8} {\bibfield  {journal} {\bibinfo  {journal} {Nature Communications}\ }\textbf {\bibinfo {volume} {16}},\ \bibinfo {pages} {7443} (\bibinfo {year} {2025})}\BibitemShut {NoStop}%
\bibitem [{\citenamefont {Ayaz}\ \emph {et~al.}(2022)\citenamefont {Ayaz}, \citenamefont {Scalfi}, \citenamefont {Dalton},\ and\ \citenamefont {Netz}}]{Ayaz2022}%
  \BibitemOpen
  \bibfield  {author} {\bibinfo {author} {\bibfnamefont {C.}~\bibnamefont {Ayaz}}, \bibinfo {author} {\bibfnamefont {L.}~\bibnamefont {Scalfi}}, \bibinfo {author} {\bibfnamefont {B.~A.}\ \bibnamefont {Dalton}}, \ and\ \bibinfo {author} {\bibfnamefont {R.~R.}\ \bibnamefont {Netz}},\ }\bibfield  {title} {\enquote {\bibinfo {title} {Generalized langevin equation with a nonlinear potential of mean force and nonlinear memory friction from a hybrid projection scheme},}\ }\href {\doibase 10.1103/PhysRevE.105.054138} {\bibfield  {journal} {\bibinfo  {journal} {Physical Review E}\ }\textbf {\bibinfo {volume} {105}},\ \bibinfo {pages} {054138} (\bibinfo {year} {2022})}\BibitemShut {NoStop}%
\bibitem [{\citenamefont {Dalton}\ \emph {et~al.}(2025)\citenamefont {Dalton}, \citenamefont {Klimek}, \citenamefont {Kiefer}, \citenamefont {Brünig}, \citenamefont {Colinet}, \citenamefont {Tepper}, \citenamefont {Abbasi},\ and\ \citenamefont {Netz}}]{Dalton2025}%
  \BibitemOpen
  \bibfield  {author} {\bibinfo {author} {\bibfnamefont {B.~A.}\ \bibnamefont {Dalton}}, \bibinfo {author} {\bibfnamefont {A.}~\bibnamefont {Klimek}}, \bibinfo {author} {\bibfnamefont {H.}~\bibnamefont {Kiefer}}, \bibinfo {author} {\bibfnamefont {F.~N.}\ \bibnamefont {Brünig}}, \bibinfo {author} {\bibfnamefont {H.}~\bibnamefont {Colinet}}, \bibinfo {author} {\bibfnamefont {L.}~\bibnamefont {Tepper}}, \bibinfo {author} {\bibfnamefont {A.}~\bibnamefont {Abbasi}}, \ and\ \bibinfo {author} {\bibfnamefont {R.~R.}\ \bibnamefont {Netz}},\ }\bibfield  {title} {\enquote {\bibinfo {title} {Memory and friction: From the nanoscale to the macroscale},}\ }\href {\doibase 10.1146/ANNUREV-PHYSCHEM-082423-031037} {\bibfield  {journal} {\bibinfo  {journal} {Annual Review of Physical Chemistry}\ }\textbf {\bibinfo {volume} {76}},\ \bibinfo {pages} {431--454} (\bibinfo {year} {2025})}\BibitemShut {NoStop}%
\bibitem [{\citenamefont {Klimek}, \citenamefont {Dalton},\ and\ \citenamefont {Netz}(2025)}]{Klimek2025}%
  \BibitemOpen
  \bibfield  {author} {\bibinfo {author} {\bibfnamefont {A.}~\bibnamefont {Klimek}}, \bibinfo {author} {\bibfnamefont {B.~A.}\ \bibnamefont {Dalton}}, \ and\ \bibinfo {author} {\bibfnamefont {R.~R.}\ \bibnamefont {Netz}},\ }\bibfield  {title} {\enquote {\bibinfo {title} {Subdiffusion from competition between multi-exponential friction memory and energy barriers},}\ }\href {\doibase 10.1140/EPJE/S10189-025-00518-Y} {\bibfield  {journal} {\bibinfo  {journal} {The European Physical Journal E 2025 48:8}\ }\textbf {\bibinfo {volume} {48}},\ \bibinfo {pages} {1--11} (\bibinfo {year} {2025})}\BibitemShut {NoStop}%
\bibitem [{\citenamefont {Hummer}\ and\ \citenamefont {Szabo}(2014)}]{Hummer2014}%
  \BibitemOpen
  \bibfield  {author} {\bibinfo {author} {\bibfnamefont {G.}~\bibnamefont {Hummer}}\ and\ \bibinfo {author} {\bibfnamefont {A.}~\bibnamefont {Szabo}},\ }\bibfield  {title} {\enquote {\bibinfo {title} {Optimal dimensionality reduction of multistate kinetic and markov-state models},}\ }\href {\doibase 10.1021/JP508375Q} {\bibfield  {journal} {\bibinfo  {journal} {Journal of Physical Chemistry B}\ }\textbf {\bibinfo {volume} {119}},\ \bibinfo {pages} {9029--9037} (\bibinfo {year} {2014})}\BibitemShut {NoStop}%
\bibitem [{\citenamefont {Forster}(1990)}]{BookForster}%
  \BibitemOpen
  \bibfield  {author} {\bibinfo {author} {\bibfnamefont {D.}~\bibnamefont {Forster}},\ }\href {\doibase 10.1201/9780429493683} {\emph {\bibinfo {title} {Hydrodynamic fluctuations, broken symmetry, and correlation functions}}}\ (\bibinfo  {publisher} {CRC Press},\ \bibinfo {year} {1990})\ pp.\ \bibinfo {pages} {1--326}\BibitemShut {NoStop}%
\bibitem [{Note6()}]{Note6}%
  \BibitemOpen
  \bibinfo {note} {Analytically, taking time derivatives of the current begins to include operators describing the phonon bath, and is therefore anything but simple (even unobtainable from the point of view of the HEOM solver). However sufficiently high time resolution allows a numerical estimate.}\BibitemShut {Stop}%
\bibitem [{Note7()}]{Note7}%
  \BibitemOpen
  \bibinfo {note} {The $\langle i,j\rangle $ notation means when this value is greater than 5~\r A~we zero the element, as described in the Methods section}\BibitemShut {NoStop}%
\bibitem [{\citenamefont {Dominic}\ \emph {et~al.}(2023{\natexlab{b}})\citenamefont {Dominic}, \citenamefont {Cao}, \citenamefont {Montoya-Castillo},\ and\ \citenamefont {Huang}}]{Dominic2023}%
  \BibitemOpen
  \bibfield  {author} {\bibinfo {author} {\bibfnamefont {A.~J.}\ \bibnamefont {Dominic}}, \bibinfo {author} {\bibfnamefont {S.}~\bibnamefont {Cao}}, \bibinfo {author} {\bibfnamefont {A.}~\bibnamefont {Montoya-Castillo}}, \ and\ \bibinfo {author} {\bibfnamefont {X.}~\bibnamefont {Huang}},\ }\bibfield  {title} {\enquote {\bibinfo {title} {Memory unlocks the future of biomolecular dynamics: Transformative tools to uncover physical insights accurately and efficiently},}\ }\href {\doibase 10.1021/JACS.3C01095} {\bibfield  {journal} {\bibinfo  {journal} {Journal of the American Chemical Society}\ }\textbf {\bibinfo {volume} {145}},\ \bibinfo {pages} {9916--9927} (\bibinfo {year} {2023}{\natexlab{b}})}\BibitemShut {NoStop}%
\bibitem [{\citenamefont {Brotzakis}\ and\ \citenamefont {Parrinello}(2018)}]{Brotzakis2018}%
  \BibitemOpen
  \bibfield  {author} {\bibinfo {author} {\bibfnamefont {Z.~F.}\ \bibnamefont {Brotzakis}}\ and\ \bibinfo {author} {\bibfnamefont {M.}~\bibnamefont {Parrinello}},\ }\bibfield  {title} {\enquote {\bibinfo {title} {Enhanced sampling of protein conformational transitions via dynamically optimized collective variables},}\ }\href {\doibase 10.1021/ACS.JCTC.8B00827} {\bibfield  {journal} {\bibinfo  {journal} {Journal of Chemical Theory and Computation}\ }\textbf {\bibinfo {volume} {15}},\ \bibinfo {pages} {1393--1398} (\bibinfo {year} {2018})}\BibitemShut {NoStop}%
\bibitem [{\citenamefont {Mardt}\ \emph {et~al.}(2018)\citenamefont {Mardt}, \citenamefont {Pasquali}, \citenamefont {Wu},\ and\ \citenamefont {Noé}}]{Mardt2018}%
  \BibitemOpen
  \bibfield  {author} {\bibinfo {author} {\bibfnamefont {A.}~\bibnamefont {Mardt}}, \bibinfo {author} {\bibfnamefont {L.}~\bibnamefont {Pasquali}}, \bibinfo {author} {\bibfnamefont {H.}~\bibnamefont {Wu}}, \ and\ \bibinfo {author} {\bibfnamefont {F.}~\bibnamefont {Noé}},\ }\bibfield  {title} {\enquote {\bibinfo {title} {Vampnets for deep learning of molecular kinetics},}\ }\href {\doibase 10.1038/S41467-017-02388-1} {\bibfield  {journal} {\bibinfo  {journal} {Nature Communications}\ }\textbf {\bibinfo {volume} {9}},\ \bibinfo {pages} {1--11} (\bibinfo {year} {2018})}\BibitemShut {NoStop}%
\bibitem [{\citenamefont {Chennakesavalu}, \citenamefont {Toomer},\ and\ \citenamefont {Rotskoff}(2023)}]{Chennakesavalu2023}%
  \BibitemOpen
  \bibfield  {author} {\bibinfo {author} {\bibfnamefont {S.}~\bibnamefont {Chennakesavalu}}, \bibinfo {author} {\bibfnamefont {D.~J.}\ \bibnamefont {Toomer}}, \ and\ \bibinfo {author} {\bibfnamefont {G.~M.}\ \bibnamefont {Rotskoff}},\ }\bibfield  {title} {\enquote {\bibinfo {title} {Ensuring thermodynamic consistency with invertible coarse-graining},}\ }\href {\doibase 10.1063/5.0141888} {\bibfield  {journal} {\bibinfo  {journal} {The Journal of Chemical Physics}\ }\textbf {\bibinfo {volume} {158}},\ \bibinfo {pages} {124126} (\bibinfo {year} {2023})}\BibitemShut {NoStop}%
\bibitem [{\citenamefont {Devergne}\ \emph {et~al.}(2025)\citenamefont {Devergne}, \citenamefont {Kostic}, \citenamefont {Pontil},\ and\ \citenamefont {Parrinello}}]{Devergne2025}%
  \BibitemOpen
  \bibfield  {author} {\bibinfo {author} {\bibfnamefont {T.}~\bibnamefont {Devergne}}, \bibinfo {author} {\bibfnamefont {V.}~\bibnamefont {Kostic}}, \bibinfo {author} {\bibfnamefont {M.}~\bibnamefont {Pontil}}, \ and\ \bibinfo {author} {\bibfnamefont {M.}~\bibnamefont {Parrinello}},\ }\bibfield  {title} {\enquote {\bibinfo {title} {The seeds of the future are in the present: A blind exploration of metastable states},}\ }\href {https://arxiv.org/pdf/2508.01477} {\bibfield  {journal} {\bibinfo  {journal} {arXiv:2508.01477 [physics.chem-ph]}\ } (\bibinfo {year} {2025})}\BibitemShut {NoStop}%
\bibitem [{\citenamefont {Ye}\ \emph {et~al.}(2019)\citenamefont {Ye}, \citenamefont {Hu}, \citenamefont {Li}, \citenamefont {Zhang}, \citenamefont {Zhong}, \citenamefont {Zhang}, \citenamefont {Luo}, \citenamefont {Mukamel},\ and\ \citenamefont {Jiang}}]{Ye2019}%
  \BibitemOpen
  \bibfield  {author} {\bibinfo {author} {\bibfnamefont {S.}~\bibnamefont {Ye}}, \bibinfo {author} {\bibfnamefont {W.}~\bibnamefont {Hu}}, \bibinfo {author} {\bibfnamefont {X.}~\bibnamefont {Li}}, \bibinfo {author} {\bibfnamefont {J.}~\bibnamefont {Zhang}}, \bibinfo {author} {\bibfnamefont {K.}~\bibnamefont {Zhong}}, \bibinfo {author} {\bibfnamefont {G.}~\bibnamefont {Zhang}}, \bibinfo {author} {\bibfnamefont {Y.}~\bibnamefont {Luo}}, \bibinfo {author} {\bibfnamefont {S.}~\bibnamefont {Mukamel}}, \ and\ \bibinfo {author} {\bibfnamefont {J.}~\bibnamefont {Jiang}},\ }\bibfield  {title} {\enquote {\bibinfo {title} {A neural network protocol for electronic excitations of n-methylacetamide},}\ }\href {\doibase 10.1073/PNAS.1821044116} {\bibfield  {journal} {\bibinfo  {journal} {Proceedings of the National Academy of Sciences}\ }\textbf {\bibinfo {volume} {116}},\ \bibinfo {pages} {11612--11617} (\bibinfo {year} {2019})}\BibitemShut {NoStop}%
\bibitem [{\citenamefont {Chen}\ \emph {et~al.}(2020)\citenamefont {Chen}, \citenamefont {Zuehlsdorff}, \citenamefont {Morawietz}, \citenamefont {Isborn},\ and\ \citenamefont {Markland}}]{Chen2020}%
  \BibitemOpen
  \bibfield  {author} {\bibinfo {author} {\bibfnamefont {M.~S.}\ \bibnamefont {Chen}}, \bibinfo {author} {\bibfnamefont {T.~J.}\ \bibnamefont {Zuehlsdorff}}, \bibinfo {author} {\bibfnamefont {T.}~\bibnamefont {Morawietz}}, \bibinfo {author} {\bibfnamefont {C.~M.}\ \bibnamefont {Isborn}}, \ and\ \bibinfo {author} {\bibfnamefont {T.~E.}\ \bibnamefont {Markland}},\ }\bibfield  {title} {\enquote {\bibinfo {title} {Exploiting machine learning to efficiently predict multidimensional optical spectra in complex environments},}\ }\href {\doibase 10.1021/ACS.JPCLETT.0C02168} {\bibfield  {journal} {\bibinfo  {journal} {The Journal of Physical Chemistry Letters}\ }\textbf {\bibinfo {volume} {11}},\ \bibinfo {pages} {7559--7568} (\bibinfo {year} {2020})}\BibitemShut {NoStop}%
\bibitem [{\citenamefont {Farahvash}\ \emph {et~al.}(2020)\citenamefont {Farahvash}, \citenamefont {Lee}, \citenamefont {Sun}, \citenamefont {Shi},\ and\ \citenamefont {Willard}}]{Farahvash2020}%
  \BibitemOpen
  \bibfield  {author} {\bibinfo {author} {\bibfnamefont {A.}~\bibnamefont {Farahvash}}, \bibinfo {author} {\bibfnamefont {C.~K.}\ \bibnamefont {Lee}}, \bibinfo {author} {\bibfnamefont {Q.}~\bibnamefont {Sun}}, \bibinfo {author} {\bibfnamefont {L.}~\bibnamefont {Shi}}, \ and\ \bibinfo {author} {\bibfnamefont {A.~P.}\ \bibnamefont {Willard}},\ }\bibfield  {title} {\enquote {\bibinfo {title} {Machine learning frenkel hamiltonian parameters to accelerate simulations of exciton dynamics},}\ }\href {\doibase 10.1063/5.0016009} {\bibfield  {journal} {\bibinfo  {journal} {The Journal of Chemical Physics}\ }\textbf {\bibinfo {volume} {153}},\ \bibinfo {pages} {74111} (\bibinfo {year} {2020})}\BibitemShut {NoStop}%
\bibitem [{\citenamefont {Kelly}\ \emph {et~al.}(2025)\citenamefont {Kelly}, \citenamefont {Hu}, \citenamefont {Damiani}, \citenamefont {Chen}, \citenamefont {Snider}, \citenamefont {Son}, \citenamefont {Lee}, \citenamefont {Gupta}, \citenamefont {Montoya-Castillo}, \citenamefont {Zuehlsdorff}, \citenamefont {Schlau-Cohen}, \citenamefont {Isborn},\ and\ \citenamefont {Markland}}]{Kelly2025}%
  \BibitemOpen
  \bibfield  {author} {\bibinfo {author} {\bibfnamefont {J.}~\bibnamefont {Kelly}}, \bibinfo {author} {\bibfnamefont {F.}~\bibnamefont {Hu}}, \bibinfo {author} {\bibfnamefont {A.}~\bibnamefont {Damiani}}, \bibinfo {author} {\bibfnamefont {M.~S.}\ \bibnamefont {Chen}}, \bibinfo {author} {\bibfnamefont {A.}~\bibnamefont {Snider}}, \bibinfo {author} {\bibfnamefont {M.}~\bibnamefont {Son}}, \bibinfo {author} {\bibfnamefont {A.}~\bibnamefont {Lee}}, \bibinfo {author} {\bibfnamefont {P.}~\bibnamefont {Gupta}}, \bibinfo {author} {\bibfnamefont {A.}~\bibnamefont {Montoya-Castillo}}, \bibinfo {author} {\bibfnamefont {T.~J.}\ \bibnamefont {Zuehlsdorff}}, \bibinfo {author} {\bibfnamefont {G.~S.}\ \bibnamefont {Schlau-Cohen}}, \bibinfo {author} {\bibfnamefont {C.~M.}\ \bibnamefont {Isborn}}, \ and\ \bibinfo {author} {\bibfnamefont {T.~E.}\ \bibnamefont {Markland}},\ }\bibfield  {title} {\enquote {\bibinfo {title} {Two-dimensional electronic spectroscopy in the condensed phase using equivariant transformer accelerated
  molecular dynamics simulations},}\ }\href {\doibase 10.1021/ACS.JPCLETT.5C00911} {\bibfield  {journal} {\bibinfo  {journal} {The Journal of Physical Chemistry Letters}\ }\textbf {\bibinfo {volume} {16}},\ \bibinfo {pages} {5561--5569} (\bibinfo {year} {2025})}\BibitemShut {NoStop}%
\bibitem [{\citenamefont {Janssen}(2018)}]{Janssen2018}%
  \BibitemOpen
  \bibfield  {author} {\bibinfo {author} {\bibfnamefont {L.~M.}\ \bibnamefont {Janssen}},\ }\bibfield  {title} {\enquote {\bibinfo {title} {Mode-coupling theory of the glass transition: A primer},}\ }\href {\doibase 10.3389/FPHY.2018.00097} {\bibfield  {journal} {\bibinfo  {journal} {Frontiers in Physics}\ }\textbf {\bibinfo {volume} {6}},\ \bibinfo {pages} {403911} (\bibinfo {year} {2018})}\BibitemShut {NoStop}%
\bibitem [{\citenamefont {Jung}, \citenamefont {Videla},\ and\ \citenamefont {Batista}(2018)}]{Jung2018}%
  \BibitemOpen
  \bibfield  {author} {\bibinfo {author} {\bibfnamefont {K.~A.}\ \bibnamefont {Jung}}, \bibinfo {author} {\bibfnamefont {P.~E.}\ \bibnamefont {Videla}}, \ and\ \bibinfo {author} {\bibfnamefont {V.~S.}\ \bibnamefont {Batista}},\ }\bibfield  {title} {\enquote {\bibinfo {title} {Inclusion of nuclear quantum effects for simulations of nonlinear spectroscopy},}\ }\href {\doibase 10.1063/1.5036768} {\bibfield  {journal} {\bibinfo  {journal} {The Journal of Chemical Physics}\ }\textbf {\bibinfo {volume} {148}},\ \bibinfo {pages} {244105} (\bibinfo {year} {2018})}\BibitemShut {NoStop}%
\bibitem [{\citenamefont {Mukamel}(1999)}]{BookMukamel}%
  \BibitemOpen
  \bibfield  {author} {\bibinfo {author} {\bibfnamefont {S.}~\bibnamefont {Mukamel}},\ }\href@noop {} {\emph {\bibinfo {title} {Principles of Nonlinear Optical Spectroscopy}}}\ (\bibinfo  {publisher} {Oxford University Press},\ \bibinfo {year} {1999})\BibitemShut {NoStop}%
\bibitem [{\citenamefont {Pouthier}, \citenamefont {Hoang},\ and\ \citenamefont {Girardet}(1999)}]{Pouthier1999}%
  \BibitemOpen
  \bibfield  {author} {\bibinfo {author} {\bibfnamefont {V.}~\bibnamefont {Pouthier}}, \bibinfo {author} {\bibfnamefont {P.~N.}\ \bibnamefont {Hoang}}, \ and\ \bibinfo {author} {\bibfnamefont {C.}~\bibnamefont {Girardet}},\ }\bibfield  {title} {\enquote {\bibinfo {title} {Infrared and infrared-visible sum frequency generation spectroscopic response of harmonic monolayer vibrons: Homogeneous profile},}\ }\href {\doibase 10.1063/1.478602} {\bibfield  {journal} {\bibinfo  {journal} {The Journal of Chemical Physics}\ }\textbf {\bibinfo {volume} {110}},\ \bibinfo {pages} {6963--6976} (\bibinfo {year} {1999})}\BibitemShut {NoStop}%
\bibitem [{\citenamefont {Bader}\ and\ \citenamefont {Berne}(1994)}]{Bader1994}%
  \BibitemOpen
  \bibfield  {author} {\bibinfo {author} {\bibfnamefont {J.~S.}\ \bibnamefont {Bader}}\ and\ \bibinfo {author} {\bibfnamefont {B.~J.}\ \bibnamefont {Berne}},\ }\bibfield  {title} {\enquote {\bibinfo {title} {Quantum and classical relaxation rates from classical simulations},}\ }\href {\doibase 10.1063/1.466780} {\bibfield  {journal} {\bibinfo  {journal} {The Journal of Chemical Physics}\ }\textbf {\bibinfo {volume} {100}},\ \bibinfo {pages} {8359--8366} (\bibinfo {year} {1994})}\BibitemShut {NoStop}%
\bibitem [{\citenamefont {Mahan}(2000)}]{BookMahan}%
  \BibitemOpen
  \bibfield  {author} {\bibinfo {author} {\bibfnamefont {G.~D.}\ \bibnamefont {Mahan}},\ }\href {\doibase 10.1007/978-1-4757-5714-9} {\emph {\bibinfo {title} {Many-Particle Physics}}}\ (\bibinfo  {publisher} {Springer US},\ \bibinfo {year} {2000})\ pp.\ \bibinfo {pages} {621--626}\BibitemShut {NoStop}%
\bibitem [{\citenamefont {Ramírez}\ \emph {et~al.}(2004)\citenamefont {Ramírez}, \citenamefont {López-Ciudad}, \citenamefont {P},\ and\ \citenamefont {Marx}}]{Ramirez2004}%
  \BibitemOpen
  \bibfield  {author} {\bibinfo {author} {\bibfnamefont {R.}~\bibnamefont {Ramírez}}, \bibinfo {author} {\bibfnamefont {T.}~\bibnamefont {López-Ciudad}}, \bibinfo {author} {\bibfnamefont {P.~K.}\ \bibnamefont {P}}, \ and\ \bibinfo {author} {\bibfnamefont {D.}~\bibnamefont {Marx}},\ }\bibfield  {title} {\enquote {\bibinfo {title} {Quantum corrections to classical time-correlation functions: Hydrogen bonding and anharmonic floppy modes},}\ }\href {\doibase 10.1063/1.1774986} {\bibfield  {journal} {\bibinfo  {journal} {The Journal of Chemical Physics}\ }\textbf {\bibinfo {volume} {121}},\ \bibinfo {pages} {3973--3983} (\bibinfo {year} {2004})}\BibitemShut {NoStop}%
\bibitem [{Note8()}]{Note8}%
  \BibitemOpen
  \bibinfo {note} {Confirmed in a private communication with the authors.}\BibitemShut {Stop}%
\bibitem [{\citenamefont {Reichman}\ \emph {et~al.}(2000)\citenamefont {Reichman}, \citenamefont {Roy}, \citenamefont {Jang},\ and\ \citenamefont {Voth}}]{Reichman2000}%
  \BibitemOpen
  \bibfield  {author} {\bibinfo {author} {\bibfnamefont {D.~R.}\ \bibnamefont {Reichman}}, \bibinfo {author} {\bibfnamefont {P.~N.}\ \bibnamefont {Roy}}, \bibinfo {author} {\bibfnamefont {S.}~\bibnamefont {Jang}}, \ and\ \bibinfo {author} {\bibfnamefont {G.~A.}\ \bibnamefont {Voth}},\ }\bibfield  {title} {\enquote {\bibinfo {title} {A feynman path centroid dynamics approach for the computation of time correlation functions involving nonlinear operators},}\ }\href {\doibase 10.1063/1.481872} {\bibfield  {journal} {\bibinfo  {journal} {The Journal of Chemical Physics}\ }\textbf {\bibinfo {volume} {113}},\ \bibinfo {pages} {919--929} (\bibinfo {year} {2000})}\BibitemShut {NoStop}%
\bibitem [{\citenamefont {Jung}\ and\ \citenamefont {Markland}(2022)}]{Jung2022}%
  \BibitemOpen
  \bibfield  {author} {\bibinfo {author} {\bibfnamefont {K.~A.}\ \bibnamefont {Jung}}\ and\ \bibinfo {author} {\bibfnamefont {T.~E.}\ \bibnamefont {Markland}},\ }\bibfield  {title} {\enquote {\bibinfo {title} {2d spectroscopies from condensed phase dynamics: Accessing third-order response properties from equilibrium multi-time correlation functions},}\ }\href {\doibase 10.1063/5.0107087} {\bibfield  {journal} {\bibinfo  {journal} {The Journal of Chemical Physics}\ }\textbf {\bibinfo {volume} {157}} (\bibinfo {year} {2022}),\ 10.1063/5.0107087}\BibitemShut {NoStop}%
\bibitem [{\citenamefont {DeVane}\ \emph {et~al.}(2004)\citenamefont {DeVane}, \citenamefont {Space}, \citenamefont {Perry}, \citenamefont {Neipert}, \citenamefont {Ridley},\ and\ \citenamefont {Keyes}}]{DeVane2004}%
  \BibitemOpen
  \bibfield  {author} {\bibinfo {author} {\bibfnamefont {R.}~\bibnamefont {DeVane}}, \bibinfo {author} {\bibfnamefont {B.}~\bibnamefont {Space}}, \bibinfo {author} {\bibfnamefont {A.}~\bibnamefont {Perry}}, \bibinfo {author} {\bibfnamefont {C.}~\bibnamefont {Neipert}}, \bibinfo {author} {\bibfnamefont {C.}~\bibnamefont {Ridley}}, \ and\ \bibinfo {author} {\bibfnamefont {T.}~\bibnamefont {Keyes}},\ }\bibfield  {title} {\enquote {\bibinfo {title} {A time correlation function theory of two-dimensional infrared spectroscopy with applications to liquid water},}\ }\href {\doibase 10.1063/1.1776119} {\bibfield  {journal} {\bibinfo  {journal} {The Journal of Chemical Physics}\ }\textbf {\bibinfo {volume} {121}},\ \bibinfo {pages} {3688--3701} (\bibinfo {year} {2004})}\BibitemShut {NoStop}%
\bibitem [{\citenamefont {Resta}(1998)}]{Resta1998}%
  \BibitemOpen
  \bibfield  {author} {\bibinfo {author} {\bibfnamefont {R.}~\bibnamefont {Resta}},\ }\bibfield  {title} {\enquote {\bibinfo {title} {Quantum-mechanical position operator in extended systems},}\ }\href {\doibase 10.1103/PhysRevLett.80.1800} {\bibfield  {journal} {\bibinfo  {journal} {Physical Review Letters}\ }\textbf {\bibinfo {volume} {80}},\ \bibinfo {pages} {1800--1803} (\bibinfo {year} {1998})},\ \bibinfo {note} {chao says this is a key paper in modern theory of P}\BibitemShut {NoStop}%
\bibitem [{\citenamefont {Spaldin}(2012)}]{Spaldin2012}%
  \BibitemOpen
  \bibfield  {author} {\bibinfo {author} {\bibfnamefont {N.~A.}\ \bibnamefont {Spaldin}},\ }\bibfield  {title} {\enquote {\bibinfo {title} {A beginner's guide to the modern theory of polarization},}\ }\href {\doibase 10.1016/J.JSSC.2012.05.010} {\bibfield  {journal} {\bibinfo  {journal} {Journal of Solid State Chemistry}\ }\textbf {\bibinfo {volume} {195}},\ \bibinfo {pages} {2--10} (\bibinfo {year} {2012})}\BibitemShut {NoStop}%
\bibitem [{\citenamefont {Schmid}(2021)}]{Schmid2022}%
  \BibitemOpen
  \bibfield  {author} {\bibinfo {author} {\bibfnamefont {P.~J.}\ \bibnamefont {Schmid}},\ }\bibfield  {title} {\enquote {\bibinfo {title} {Dynamic mode decomposition and its variants},}\ }\href {\doibase 10.1146/ANNUREV-FLUID-030121-015835} {\bibfield  {journal} {\bibinfo  {journal} {Annual Review of Fluid Mechanics}\ }\textbf {\bibinfo {volume} {54}},\ \bibinfo {pages} {225--254} (\bibinfo {year} {2021})}\BibitemShut {NoStop}%
\end{thebibliography}%

\end{document}